\documentclass[11pt]{article}

\usepackage{amsmath}
\usepackage{amssymb}
\usepackage{graphicx}
\usepackage[T2A,T1]{fontenc}
\usepackage{jheppub}
\usepackage{xcolor}
\usepackage{soul}

\hypersetup{
  colorlinks=true,
  citecolor=blue,
  linkcolor=black,
  urlcolor=blue,
  linktocpage,
  linktoc=all
}

\numberwithin{equation}{section}
\newcommand{\ft}[2]{{\textstyle{\frac{#1}{#2}}}}

\def\q{{\mathfrak{q}}}
\def\w{{\mathfrak{w}}}

\def\calo{{{\cal O} }}
\def\calf{{{\cal F} }}

\DeclareRobustCommand{\dje}{%
  \ifmmode
    \text{{\fontencoding{T2A}\selectfont\cyrdje}}%
  \else
    {\fontencoding{T2A}\selectfont\cyrdje}%
  \fi
}
\renewcommand{\hbar}{\dje} 

\begin{document}

\title{Causality violation in the holographic approach to ${\cal N}=4$ SYM theory at finite 't Hooft coupling}
\author[a]{Alex Buchel}
\author[b,c]{and Pavel Kovtun}
\affiliation[a]{
{\it \small Department of Physics and Astronomy, University of Western Ontario}\\
{\it \small London, Ontario N6A 5B7, Canada}
}
\affiliation[b]{
{\it\small School of Mathematical Sciences \& STAG Research Centre}\\
{\it\small University of Southampton, SO17 1BJ, Southampton, United Kingdom}
}
\affiliation[c]{
{\it\small Department of Physics \& Astronomy, University of Victoria}\\
{\it\small PO Box 1700 STN CSC, Victoria, BC, V8W 2Y2, Canada}
}
\abstract{
In the AdS/CFT correspondence, correlation functions of the energy-momentum tensor in ${\cal N}=4$ supersymmetric Yang-Mills theory at large but finite 't~Hooft coupling can be studied using higher-derivative corrections to supergravity. We show that the commonly used holographic prescription for correlation functions which is obtained by truncating the equations of motion at order $(\alpha')^3$ implies that the thermal equilibrium state of ${\cal N}=4$ SYM theory in the planar limit is dynamically unstable at arbitrarily large finite 't Hooft coupling. The instability is a reflection of the unphysical causality violation brought about by treating perturbatively-derived equations as if they were non-perturbative, analogous to the instability of AdS black branes in five-dimensional Gauss-Bonnet gravity. While other holographic prescriptions for treating the $O(\alpha'^{\,3})$ corrections to supergravity may alleviate the pathology introduced by the truncation, the validity regime of the results obtained by any prescription at $O(\alpha'^{\,3})$ must be limited to the strict perturbative treatment of the $O(\alpha'^{\,3})$ effects on ${\cal N}=4$ SYM observables. 
}

\maketitle

\section{Introduction}
\label{sec:introduction}

Gauge/gravity duality provides a particularly useful setting for studying real-time dynamics of strongly coupled gauge theories at finite temperature~\cite{Ammon:2015wua}. In the planar large-$N_c$ limit, and in the limit of infinite 't~Hooft coupling $\lambda$, the holographic description of $\mathcal N=4$ supersymmetric Yang-Mills (SYM) theory reduces to two-derivative supergravity, and its transport properties and quasinormal-mode (QNM) spectra are correspondingly tractable.  The effects of finite 't~Hooft coupling are encoded in higher-derivative string corrections to two-derivative supergravity. For $\mathcal N=4$ SYM theory, the leading correction is the type-IIB order $(\alpha')^3$ interaction, whose dimensionless strength is $\gamma/L^6=\zeta(3)/(8\lambda^{3/2}) > 0$, where $L$ is the radius of $AdS_5$, determined by the 5-form flux. The $O(\alpha'^{\,3})$-corrected theory has been used extensively to study finite-coupling corrections to equilibrium observables of $\mathcal N=4$ SYM theory, its transport coefficients, and QNM spectra, see e.g.~\cite{Gubser:1998nz, Buchel:2004di, Benincasa:2005qc, Buchel:2008sh, Myers:2008yi, Stricker:2013lma, Waeber:2015oka, Grozdanov:2016vgg, Casalderrey-Solana:2018rle, Grozdanov:2018kkt, Grozdanov:2021jfw} and references therein. Perturbative treatment of finite-$\lambda$ corrections in the planar limit corresponds to $\gamma\to 0$.

There is an important distinction between using the higher-derivative expansion perturbatively, and treating a truncated fluctuation equation in the bulk as an exact equation at finite non-zero $\gamma$. In many applications, the $O(\gamma)$-truncated equations, after perturbative order-reduction, are solved at finite~$\gamma$~\cite{Waeber:2015oka, Casalderrey-Solana:2018rle, Grozdanov:2021jfw}.  This procedure of treating the perturbative in~$\gamma$ second-order equations as if they were non-perturbative in $\gamma$ amounts to ``resumming'' a particular subset of formally higher-order contributions, but the result is prescription-dependent, once effects beyond the controlled $O(\gamma)$ order become important~\cite{Waeber:2015oka, Buchel:2018eax}.  At fixed frequency and momentum of QNMs, this difference between prescriptions disappears as $\gamma\to0$. However, it need not disappear uniformly when frequency or momentum are taken large at the same time.  

In this paper, we will examine precisely such a regime for linearized perturbations of D3-branes, when the perturbation equations are reduced to second order in derivatives and truncated at $O(\gamma)$. We will see that for any small value of $\gamma$, there is a value of momentum for which the corresponding large-momentum quasinormal modes violate the constraint of relativistic causality on the retarded function, the analyticity at ${\rm Im}(\omega) > |{\rm Im}(q)|$. Here $\omega$ is the frequency, and $q$ is the momentum of the perturbation. Consequently, there is a reference frame in which the system is unstable to spatially homogeneous perturbations, at a value of frequency which can be made arbitrarily small by a choice of the boost velocity. 

The situation we encounter here is analogous to the instability of black branes in five-dimensional Gauss-Bonnet (GB) gravity~\cite{Buchel:2026rep}. The fact that the five-dimensional GB theory in AdS may violate boundary causality has been known for some time \cite{Brigante:2007nu, Brigante:2008gz, Buchel:2009tt}, as has the need for an ultraviolet completion~\cite{Camanho:2014apa}; however the causality-violating quasinormal modes were only explicitly constructed recently~\cite{Buchel:2026rep}. Given the wide use of the $(\alpha')^3$-truncated theory in applied holographic studies, one may wonder whether a similar causality violation is also exhibited by the truncated bulk theory intended to describe ${\cal N}=4$ SYM at large but finite 't~Hooft coupling.  We find that the answer is affirmative, with the important difference that in the present case the boundary causality violation (and the associated instability) develops at any finite value of $\gamma$, however small, while in the GB case, theories with sufficiently small values of the higher-derivative coupling coefficient did not suffer from this type of boundary causality violation.

Our starting point is the finite-$\gamma$ QNM problem in the three helicity sectors of the stress-tensor fluctuations: helicity 2 (scalar), helicity 1 (shear), and helicity 0 (sound).  Following the strategy used in the Gauss--Bonnet analysis of \cite{Buchel:2026rep}, we rewrite each fluctuation equation as a one-dimensional Schr\"odinger equation.  We define $\w\equiv \omega/(2\pi T)$, $\q\equiv q/(2\pi T)$, where $T$ is the temperature.  For purely imaginary spatial momentum~$\q$, the role of the Planck constant is played by $\hbar =1/|\mathfrak q|$, while the energy is $E=-\mathfrak w^2/\mathfrak q^2$. We then look for QNMs whose frequency $\w$ is purely imaginary, so that $E<0$. In the limit of small~$\hbar$, the ground state energy is determined by the minimum of the potential. Bound states with $E<-1$ violate analyticity of the retarded function for ${\rm Im}(\w) > |{\rm Im}(\q)|$, and become unstable quasinormal modes in a boosted reference frame. We will find that such bound states exist for arbitrarily small positive $\gamma$. The correlation of the small-$\gamma$ instability with momentum-dependence is our central result. 
Quantitatively, the three helicity channels become
universal in the correlated limit
\begin{align}
\label{eq:c-limit}
  \gamma\to0,\qquad
  |\mathfrak q|\to\infty,\qquad
  \gamma|\mathfrak q|^2=\text{fixed}.
\end{align}
We will show that to leading order in the limit of \eqref{eq:c-limit}, the three helicity channels share the attractive potential when the perturbations are written in the Schr\"odinger-equation form,
\begin{align}
\label{eq:U0}
  V(x)
  =-(1-x^2)\left(1+192\,\gamma|\mathfrak q|^2x^5\right)\,.
\end{align}
Here $x$ is the radial coordinate, related to the standard static coordinate $r$ by $x=r_0^2/r^2$, where $r_0$ is the horizon radius, and the AdS radius $L$ has been set to 1, so that $\gamma$ is dimensionless. 
Once $\gamma|\mathfrak q|^2$ exceeds a finite threshold of $25\sqrt{15}/3456 \approx 0.028$, the potential \eqref{eq:U0} necessarily supports a bound state with $E<-1$ in the limit~\eqref{eq:c-limit}. 
The onset of causality violation therefore scales as
\begin{align}
|\mathfrak q|_{\rm onset}\sim\gamma^{-1/2}
\sim\lambda^{3/4}.
\end{align}
Thus, causality violation is a large complex-momentum effect; at fixed momentum it disappears as $\gamma\to0$, while for any fixed sufficiently small positive $\gamma$, causality violation emerges when $|\q|$ is taken sufficiently large. 

We also study the QNM spectrum in the helicity-2 channel numerically. Starting from the bound-state with $E<-1$, we follow the causality-violating mode down to $\q=0$, where it becomes the extra ``non-perturbative'' mode on the imaginary $\w$-axis, identified in~\cite{Grozdanov:2016vgg}. For this non-perturbative mode, we determine its instability rate in a boosted reference frame, as a function of the boost velocity. In a moving reference frame, the causality-violating non-perturbative mode becomes a zero-momentum unstable mode; the instability rate diverges as the boost velocity approaches zero, similar to the instability of the diffusive mode in a moving reference frame. On the other hand, the instability rate becomes small as the boost velocity approaches one: while the pathology of the truncated system manifests as an ultraviolet effect in the rest frame, it becomes an infrared effect in a moving reference frame, reflecting causality violation introduced by treating the $O(\gamma)$ perturbative equations non-perturbatively. Further, we continue the Einstein-gravity QNM spectrum to finite $\gamma$ and then into complex momentum. We find that one of the ``Christmas-tree'' branches becomes causality-violating, analogously to what happens in the Gauss-Bonnet theory~\cite{Buchel:2026rep}. The causality-violating ``Christmas-tree'' branch connects continuously to the causality-violating non-perturbative mode on the imaginary $\w$-axis.  

What about higher orders in the derivative expansion of the bulk equations of motion? To that effect, we study, in the helicity-2 channel, a phenomenological extension of the five-dimensional action through order $\gamma^2$ \cite{Buchel:2018eax}.  At the order common to the two calculations, the resulting Schr\"odinger potential agrees exactly with that obtained from the ten-dimensional $O(\gamma)$ theory.  The new momentum-enhanced $O(\gamma^2)$ contribution survives in the same correlated limit and is attractive; consequently the bound state persists for every fixed finite coefficient of the added $O(\gamma^2)$ interaction.  This is a nontrivial robustness test of the instability within that extended truncation.

It is not, however, a proof that the instability is physical and is in fact present in the complete finite-$\gamma$ theory.  The same correlated limit which produces the universal potential also enhances higher-derivative operators: contributions schematically of order $\gamma^n|\mathfrak q|^{2n}$ need not be small when $\gamma|\mathfrak q|^2$ is held fixed.  We therefore have no argument that still higher terms in the string effective action could not qualitatively modify, or even eliminate, the instability. Our results should consequently be viewed as establishing large-$\q$ instabilities of particular finite-coupling equations, while simultaneously exposing the limitations of using the leading holographic finite-coupling correction as a controlled approximation for ``non-perturbative'' modes.

The remainder of the paper is organized as follows.  We first review the holographic setup and the $O(\gamma)$-corrected black-brane background in Section~\ref{sec:holographic_setup}. We formulate the QNM problem in the three helicity channels, and map it to Schr\"odinger form. In Section~\ref{sec:QNM}, we derive the helicity-independent, small-$\gamma$ bound-states. For the helicity-2 channel, we numerically follow the unstable branch both to $\q=0$ and to large imaginary $\q$, and connect it to the ordinary ``Christmas-tree'' QNM spectrum at complex momentum.  In Section~\ref{sec:gamma2_extension}, we then analyze a model $O(\gamma^2)$ extension as a test of truncation sensitivity, observing the robustness of the instability.  We close in Section~\ref{sec:discussion} with a discussion of the range of validity of the truncated finite-$\gamma$ description.

\section{Holographic setup}
\label{sec:holographic_setup}

We start with the tree-level type IIB low-energy effective action in ten dimensions
taking into account the 
leading-order string corrections \cite{Grisaru:1986vi,Freeman:1986zh,Park:1987jp,Gross:1986iv}
\begin{equation}
S_{10}=  \frac{1}{ 16\pi G_{10}} \int d^{10} x \sqrt {-g}
\ \bigg[ R - {\frac 12} (\partial \phi)^2 - \frac{1}{4 \cdot 5!}   (F_5)^2  +...+ 
\  \gamma \ e^{- {\frac 3 2} \phi}  W + ...\bigg]   \  ,
\label{s10}
\end{equation}
$$ \ \ \ \ \ \   
  \gamma= { \frac 18} \zeta(3)(\alpha')^3 \ , 
$$  
where 
\begin{equation}
W =  C^{hmnk} C_{pmnq} C_{h}^{\ rsp} C^{q}_{\ rsk} 
 + {\frac 12}  C^{hkmn} C_{pqmn} C_h^{\ rsp} C^{q}_{\ rsk}\  ,
\label{weyl4}
\end{equation}
and $C$ is the Weyl tensor. As in \cite{Gubser:1998nz,Buchel:2004di,Benincasa:2005qc,Buchel:2008ac,Buchel:2008sh,Buchel:2008bz,Buchel:2008kd}, the self-dual $F_5$ form does not receive order $(\alpha')^3$ corrections%
\footnote{
The five-form-dependent higher-derivative terms do not contribute to the corrected background equations for the uncharged $AdS_5$ black hole/black D3-brane near-horizon geometry to order $O(\gamma)$
\cite{Paulos:2008tn}, nor do they modify the metric perturbation equations at $O(\gamma)$~\cite{Myers:2008yi}. 
}.
In \eqref{s10} ellipses stand for other fields not essential for the present analysis, and for further higher-derivative $O(\gamma^{5/3})$ corrections to supergravity~\cite{Green:1999pu}.
The precise relation between the expansion parameters in supergravity and in ${\cal N}=4$ SYM theory is given by~\cite{Aharony:1999ti}
\begin{equation}
\lambda\equiv g_{\mathrm{YM}}^2N_c=\frac{L^4}{(\alpha')^2}\,,\qquad a=c=\frac{\pi^4 L^8}{8G_{10}}=\frac{N_c^2}{4}
\qquad \text{(planar limit)},
\label{old2.6}
\end{equation}
where $\lambda$ is the 't~Hooft coupling, and $c$, $a$ are the conformal anomaly coefficients.

\subsection{Black brane background}

We represent a ten-dimensional background geometry describing $\gamma$-corrected black D3-branes 
by the following ansatz
\begin{equation}
\begin{split}
ds_{10}^2=&g_{\mu\nu}^{(0)}\ dx^\mu dx^\nu+c_4^2\left(dS^5\right)^2\\
\equiv&-c_1^2 dt^2+c_2^2\left(dx_1^2+dx_2^2+dx_3^2\right)+c_3^2dr^2+c_4^2 d\Omega_5^2  \,,
\end{split}
\label{g10b}
\end{equation}
where $c_i=c_i(r)$ and $d\Omega_5^2$ is a metric on a round five-sphere. 
For the dilaton we assume $\phi=\phi(r)$ and for the five-form 
\begin{equation}
F_5=\calf_5+\star\calf_5\,,\qquad \calf_5=-4L^4\ dvol_{S^5}\,.
\label{5form}
\end{equation}
In \eqref{5form} the 5-form flux is chosen in such a way that $\gamma=0$ solution corresponds to $c_4=L$.
To leading order in $\gamma$, solution can be written explicitly \cite{Gubser:1998nz,Pawelczyk:1998pb}
\begin{equation}
\begin{split}
c_1=&\frac rL \left(1-\frac{r_0^4}{r^4}\right)^{1/2} e^{-\ft 53 \nu}\left(1+a+4b\right)\,,\\
c_2=&\frac rL e^{-\ft 53 \nu}\,,\\
c_3=&\frac{L}{r\left(1-\frac{r_0^4}{r^4}\right)^{1/2}}\  e^{-\ft 53 \nu}\left(1+b\right)\,,\\
c_4=&Le^{\nu}\,,
\end{split}
\label{defm}
\end{equation} 
where to order $O(\gamma)$ 
\begin{equation}
\begin{split}
a=&-\frac{\gamma}{L^6}\ \frac{15r_0^4}{2r^4}\left(25 \frac{r_0^4}{r^4}-79 \frac{r_0^8}{r^8}+25\right)\,,\\
b=&\frac{\gamma}{L^6}\ \frac{15r_0^4}{2r^4}\left(5 \frac{r_0^4}{r^4}-19 \frac{r_0^8}{r^8}+5\right)\,,\\
\nu=&\frac{\gamma}{L^6}\ \frac{15r_0^8}{32r^8}\left(1+\frac{r_0^4}{r^4}\right)\,.
\end{split}
\end{equation}
The background dilaton receives an $O(\gamma)$ correction~\cite{Gubser:1998nz}. On the gravitational QNM branch, the leading dilaton fluctuation vanishes as a consequence of the ingoing and normalizable boundary conditions. In the helicity-0 channel, where metric--dilaton mixing is allowed, the metric perturbation generally sources a dilaton fluctuation $\delta\phi=O(\gamma)$. Its contribution to the linearized metric equation, through both the dilaton stress tensor and the higher-derivative interaction, begins at $O(\gamma^2)$. Consequently, the gravitational QNM equations through $O(\gamma)$ can be derived without including the background dilaton or its induced fluctuation.

The thermal equilibrium state of the boundary field theory associated with black brane~\eqref{g10b} is characterized by the temperature $T$, the entropy density $s$, the pressure $P$ and the energy density ${\cal E}$. The entropy is the Wald entropy
of the background geometry \cite{Wald:1993nt,Dutta:2006vs}:
\begin{equation}
s=\frac{2c}{\pi}\left(\frac{r_0}{L^2}\right)^3
\left[1+60\frac{\gamma}{L^6}
  + \dots \right] \,,
\label{s}
\end{equation}
where the ``$\dots$'' stand for subleading $O(\gamma^{5/3})$ corrections. 
The temperature is proportional to the surface gravity $\kappa$ at the horizon,
\begin{equation}
T=\frac{\kappa}{2\pi}= \frac{r_0}{\pi L^2}\left[1+15\frac{\gamma}{L^6} + \dots \right]\,.
\label{tk}
\end{equation}
Implementing the standard holographic renormalization for the 1-point function of the boundary stress-energy tensor \cite{Skenderis:2002wp}, we identify
\begin{equation}
\label{eq:EPs}
{\cal E}=\frac 34 Ts \,,\qquad P=\frac 14 Ts \,,\qquad s=2\pi^2c \, T^3 \left[1+15\frac{\gamma}{L^6}
  + \dots \right]
\end{equation}
where we used  \eqref{tk} to arrive at  $s(T)$.
In what follows, we set $L=1$.

\subsection{Perturbation equations}
\label{sec:perturbation_equations}
We consider linearized perturbations of the metric~\eqref{g10b} of the form $h_{MN} = e^{-i \omega t + i q x_3}\xi_{MN}(r)$, where the wave vector is chosen along $x_3$ by isotropy. We use the gauge $h_{Mr}=0$. Rotation symmetry of the equilibrium background~\eqref{g10b} implies that the equations of motion for the fluctuations separate into three distinct channels upon linearization: a helicity-2 channel (scalar), a helicity-1 channel (shear), and a helicity-0 channel (sound), analogous to what happens for black branes in Einstein gravity~\cite{Kovtun:2005ev}. 
For all three channels, the linearized equations at $O(\gamma)$ may be written in terms of variables $Z_a(x)$ which are built from linear combinations of the components of $\xi_{MN}(r)$. The equations are of the form \cite{Kovtun:2005ev, Buchel:2004di,Benincasa:2005qc,Buchel:2008bz}
\begin{equation}
\label{eq:QNM_master_eqr}
\begin{aligned}
0 &= Z_a''(x)
+ \hat{\cal A}_a(x,\omega,q^2) Z_a'(x)
+ \hat{\cal B}_a(x,\omega,q^2) Z_a(x)\\
&\quad + \gamma\biggl[
C_a^{(4)}\frac{d^4Z_a}{dx^4}
+ C_a^{(3)}\frac{d^3Z_a}{dx^3}
+ C_a^{(2)}\frac{d^2Z_a}{dx^2}
+ C_a^{(1)}\frac{dZ_a}{dx}
+ C_a^{(0)}Z_a
\biggr]\,.
\end{aligned}
\end{equation}
where $C_a^{(n)}=C_a^{(n)}(x,\omega,q^2)$ as well, and the radial coordinate $x\in[0,1]$ is related to the radial coordinate $r$ in \eqref{g10b} as
\begin{equation}
  x=\frac{r_0^2}{r^2}\,.
  \label{rtox}
\end{equation}
The index $a$ denotes the helicity of the perturbation ($a=2$ for scalar, $a=1$ for shear, $a=0$ for sound). There is no summation over $a$ in eq.~\eqref{eq:QNM_master_eqr}. The coefficients $\hat{\mathcal A}_a$, $\hat{\mathcal B}_a$ and $C_a^{(n)}$ depend on $q^2$ by rotation invariance. For the scalar channel the coefficients $C_a^{(n)}$ were first derived in~\cite{Buchel:2004di}, and for the other channels in \cite{Benincasa:2005qc}.

We are interested in computing the quasinormal modes of the black brane background~\eqref{g10b}. The differential equations \eqref{eq:QNM_master_eqr} are fourth-order equations, while the standard procedure for computing quasinormal modes assumes a second-order differential equation, with one boundary condition imposed at the horizon, and one condition imposed at the AdS boundary. 
To order $O(\gamma)$, one can derive the equivalent linearized equations for $Z_a$ by eliminating terms proportional to $\gamma Z_a^{(n)}$, with $n\geq 2$, in \eqref{eq:QNM_master_eqr}; \textit{e.g.},
\begin{equation}
\gamma C_a^{(4)}\frac{d^4Z_a}{dx^4}
=
\gamma C_a^{(4)}\frac{d^2}{dx^2}
\biggl[
-\hat{\mathcal A}_a(x,\omega,q^2) Z_a'(x)
-\hat{\mathcal B}_a(x,\omega,q^2) Z_a(x)
\biggr]
+\mathcal{O}(\gamma^2),
\end{equation}
arriving at
\begin{equation}
\label{eq:QNM_master_eq}
0=Z_a''(x)
+\mathcal A_a(x,\omega,q^2,\gamma) Z_a'(x)
+\mathcal B_a(x,\omega,q^2,\gamma) Z_a(x)\,.
\end{equation}
The explicit expressions for $\mathcal A_a$ and $\mathcal B_a$ in \eqref{eq:QNM_master_eq} are given in Appendix~\ref{app:QNM_coeffs}.
Although \eqref{eq:QNM_master_eq} is only controlled perturbatively through order $O(\gamma)$, finite-coupling holographic studies often treat this equation as exact at numerically small but finite $\gamma$, thereby implementing a partial ``resummation'' of higher-order effects. This approach was first explored in \cite{Waeber:2015oka} and later used, for example, in \cite{Casalderrey-Solana:2018rle, Grozdanov:2021jfw}, where the ``non-perturbative'' modes were given a physical interpretation.
In what follows, we adopt the same ``resummation'' approach and analyze the spectrum of ${\cal N}=4$ SYM within this framework; we return to its limitations below%
\footnote{
We stress that eq.~\eqref{eq:QNM_master_eq} which, in the ``resummation'' prescription, is treated non-perturbatively in $\gamma$, is itself the result of a double truncation: first, neglecting $O(\gamma^{5/3})$ terms at the level of the supergravity action~\eqref{s10}, and second, neglecting the $O(\gamma^2)$ terms when the linearized perturbation equations are reduced to second order in derivatives. Unsurprisingly, the results obtained in the ``resummation'' approach are sensitive to unknown higher-order corrections in $\gamma$ \cite{Buchel:2018eax}. Consequently, any predictions of eq.~\eqref{eq:QNM_master_eq} which extend beyond the linear order in $\gamma$ are unreliable. As we will see later in the paper, the consequences of treating eq.~\eqref{eq:QNM_master_eq} non-perturbatively in $\gamma$ include results which are not only unreliable (being truncation artefacts), but are also pathological, predicting unphysical causality violations in ${\cal N} = 4$ SYM theory. 
}.

We will be primarily working with the dimensionless 
\begin{equation}
  \w = \frac{\omega}{2\pi T}\,, \qquad \q = \frac{q}{2\pi T}\,,
  \label{defwq}
\end{equation}
where $T$ is the temperature. The perturbative in $\gamma$ temperature is given by eq.~\eqref{tk}. We will find it more technically convenient to work with 
\begin{equation}
  T=\frac{r_0}{\pi\sqrt{1-30\gamma}}=\frac{r_0}{\pi} \left(1+15\gamma+ \dots \right) 
  \label{tknp}
\end{equation}
in the definitions \eqref{defwq} because this choice gives the horizon indices of the differential equation \eqref{eq:QNM_master_eq} as $\pm i\w/2$, just like in Einstein gravity \cite{Buchel:2008sh}. The choice of $T$ which is used in the normalization of $\omega$ and $q$ in \eqref{defwq} does not affect the results in the scaling limit \eqref{eq:c-limit}, as long as this $T$ agrees with the perturbative temperature~\eqref{tk} at $O(\gamma)$. In what follows, we will find it convenient to set $r_0 = 1/2$, so that $2\pi T = 1$ in the two-derivative supergravity approximation.

For quasinormal modes, we choose the infalling solution for $Z_a(x)$, whose near-horizon expansion starts with $(1-x)^{-i\w/2}$.  Near the boundary, the indices are $0$ and $2$ just like in the Einstein gravity \cite{Benincasa:2005qc}. Thus, the solution which is infalling near the horizon, will behave near the boundary as
\begin{align}
  Z_a(x) = F_{(a)}(\w, \q^2, \gamma) (1 + \dots) + G_{(a)}(\w, \q^2, \gamma) x^2 + \dots\,,
\end{align}
where the dots denote subleading terms, including logarithms. The quasinormal modes are obtained by imposing the Dirichlet condition at the boundary, thus setting
\begin{align}
\label{eq:qnmf}
  F_{(a)}(\w, \q^2, \gamma) = 0 \,.
\end{align}
Solutions of eq.~\eqref{eq:qnmf} give the singularities of the retarded functions of the energy-momentum tensor in the boundary field theory~\cite{Kovtun:2005ev}.

The differential equation \eqref{eq:QNM_master_eq} can be rewritten in the form of Schr\"odinger's equation. Similar to the analysis in \cite{Buchel:2026rep}, we define a new variable $y(x)$,
\begin{equation}
    \frac{dy}{dx} = g_a(x)\,,
\end{equation}
with $g_a(x)$ sign-definite on $(0,1)$, so that $y(x)$ is monotonic. In the resulting differential equation for $Z_a(y)$, we further take
\begin{align}
  Z_a(y) = \frac{\psi_a(y)}{{\cal C}_a(y)} \,,
\end{align}
and choose ${\cal C}_a(y)$ to eliminate the first derivative in the equation for $\psi_a(y)$,
\begin{align}
\label{eq:CC}
  \frac{{\cal C}_a'(y)}{{\cal C}_a(y)} = \frac12 \left( \frac{g_a'}{g_a^2} + \frac{{\cal A}_a}{g_a} \right) .
\end{align}
We choose the function $g_a(x)$ so that the resulting differential equation for $\psi_a(y)$, in the limit $|\q^2|\to\infty$, $|\w^2| \to\infty$, with $|\w^2|/|\q^2|$ fixed and for
$\gamma=0$, would take the form 
\begin{align}
  \psi_a''(y) + \q^2 U_a^{(0)}(y) \psi_a(y) = -\w^2 \psi_a(y) \,,
\end{align}
where the potential $U_a^{(0)}(y)$ is finite in the limit. This is achieved by
\begin{align}
g_a(x) = -\frac{1}{x^{1/2} (1-x^2)}\,,\qquad   y=-\operatorname{arctanh}\sqrt{x}-\arctan\sqrt{x}\,,
\label{eq:yx}
  \end{align}
the same function in all three symmetry channels~\cite{Buchel:2009sk}.
The relation between $x$ and $y$ near the boundary and near the horizon is
\begin{equation}
\begin{aligned}
y&=-2\sqrt{x}+O(x^{5/2}),
&& x\to0,\\
y&=\frac12\ln(1-x)-\ln2-\frac{\pi}{4}+O(1-x),
&& x\to1\,.
\end{aligned}
\label{n2.8}
\end{equation}
Given $g_a(x)$, the function ${\cal C}_a$ is obtained from \eqref{eq:CC}. We now define 
\begin{align}
  \hbar^2 \equiv -\frac{1}{\q^2}\,,\ \ \ \ E\equiv - \frac{\w^2}{\q^2}\,.
\end{align}
Perturbation equations~\eqref{eq:QNM_master_eq} then become
\begin{align}
\label{eq:Scheq}
  -\hbar^2 \frac{d^2 \psi_a}{dy^2} + U_a(y) \psi_a(y) = E \psi_a(y) \,,
\end{align}
where the effective potential $U_a(y)$ is determined by the coefficients ${\cal A}_a$ and ${\cal B}_a$ in \eqref{eq:QNM_master_eq}. We find that in all three helicity channels
\begin{equation}
  U_a=\frac{1}{\hbar^2}\ U_a^{(-2)}(y,E,\gamma,T)+ \ U_a^{(0)}(y,E,\gamma,T)+ {\hbar^2}\ U_a^{(2)}(y,E,\gamma,T)
\label{eq:uan}
  \end{equation}
where we retained the dependence on $T$ in $U_a^{(n)}$ terms above, rather than substituting the expression \eqref{tknp}: in this way $U_a^{(n)}$ are polynomials in $\gamma$, with the  $O(\gamma)$ terms being robust, while the subleading $\gamma$-terms being sensitive to higher-order corrections in \eqref{s10}. We collect the explicit leading-order expressions for $U_a^{(n)}$ in Appendix \ref{ap:us}. Note the crucial difference from the analysis in \cite{Buchel:2026rep}: in the Gauss-Bonnet holographic model all $U_a^{(-2)}$ terms were absent.

\section{Unstable quasinormal modes}
\label{sec:QNM}

\subsection{Signatures of the instability}
\label{sec:instab-2}

We closely follow the analysis of \cite{Buchel:2026rep}.
The quasinormal frequencies of the background \eqref{g10b} are determined by eq.~\eqref{eq:qnmf}. Even though it appears that eq.~\eqref{eq:qnmf} only determines the spectrum of the quasinormal modes for the black brane ``at rest'', the same equation also determines the spectrum in a uniformly boosted thermal state. Indeed, applying a Lorentz transformation to the boundary coordinates maps the equilibrium black brane to a solution characterized by a constant timelike unit four-velocity $u'^\mu$. The QNM condition in the boosted state can therefore be expressed in terms of Lorentz invariants constructed from the mode four-momentum and $u'^\mu$. 
Define the mode four-momentum in the rest frame as $Q^\mu=(\w,0,0,\q)$. Under a Lorentz boost in the $z$-direction,
\begin{align*}
Q'^\mu=\frac{1}{\sqrt{1-v^2}}
\left(\w+v\q,\,0,\,0,\,\q+v\w\right).
\end{align*}
The same boost maps the fluid four-velocity from $u^\mu=(1,0,0,0)$ to
\begin{align*}
 u'^\mu=\frac{(1,0,0,v)}{\sqrt{1-v^2}}.
\end{align*}
The QNM condition in the boosted state is therefore
\begin{align}
\label{eq:FF}
F_{(a)}(\Omega,\mathcal K^2,\gamma)=0,
\end{align}
where the Lorentz invariants are constructed from the boosted momentum and fluid velocity:
\begin{align*}
\Omega
& \equiv-Q'^\mu\eta_{\mu\nu}u'^\nu
=\frac{\w'-v\q'}{\sqrt{1-v^2}}=\w,\\
\mathcal K^2
&\equiv Q'^\mu(\eta_{\mu\nu}+u'_\mu u'_\nu)Q'^\nu
=\frac{(\q'-v\w')^2}{1-v^2}=\q^2.
\end{align*}
Thus, whenever $(\w,\q^2)$ satisfies the rest-frame QNM condition, the Lorentz-transformed pair $(\w',\q'^2)$ describes a QNM of the boosted state.

We will be looking for quasinormal modes of \eqref{g10b} which satisfy ${\rm Im}(\w) > |{\rm Im}(\q)|$ (such modes violate relativistic causality, which demands analyticity of the retarded function for ${\rm Im}(\w) > |{\rm Im}(\q)|$). We will consider modes $(\w, \q)$ which become homogeneous unstable modes after a boost, namely $(\w', \q') = (i\Gamma, 0)$, with $\Gamma>0$. Such modes are given by
\begin{align}
\label{eq:wq-2}
  (\w, \q) = \frac{i\Gamma}{\sqrt{1-v^2}} \left( 1, -v \right) ,
\end{align}
with both $\w$ and $\q$ purely imaginary, $\w^2/\q^2 = 1/v^2 > 1$, and hence for the unstable modes
\begin{equation}
\label{eq:hbarE}
    \hbar^2 =  \frac{(1-v^2)}{v^2 \Gamma^2}>0 \,, \qquad E = -\frac{1}{v^2} < -1\,.
\end{equation}
For the mode~\eqref{eq:wq-2} with $\Gamma>0$, the boundary condition on $Z_a(x)$ at the horizon becomes
\begin{equation}
    Z_a(x) \propto (1-x)^{-i\w/2} = (1-x)^{\frac{\Gamma}{2\sqrt{1-v^2}}}\,,
\end{equation}
thus $Z_a(x)$ obeys Dirichlet conditions at both the horizon $x=1$ and the boundary $x=0$. 
From Appendix \ref{app:QNM_coeffs}, in all three channels,
\begin{equation}
\label{eq:C_asymptotics}
    {\cal C}_a(x) = 
    \begin{cases}
        O(x^{-3/4}),& x\to0\,,\\[5pt]
        O(1)\,, & x \to 1 \,.
    \end{cases}
\end{equation}
As $Z_a(x) = O(x^2)$ at the boundary, $\psi_a(x)$ also obeys Dirichlet boundary conditions at both $x=0$ and $x=1$. The question of finding the modes described above is therefore mapped onto a one-dimensional quantum-mechanical problem of finding negative-energy states for a potential with infinite walls at $x=0$ and $x=1$ (rather, $y=0$ and $y=-\infty$). The unstable modes correspond to bound states with energies $E<-1$.

\subsection{Causality-violating bound states for all helicities}
\label{sec:universal_bound_states}

\begin{figure}
\centering
\includegraphics[width=0.6\textwidth]{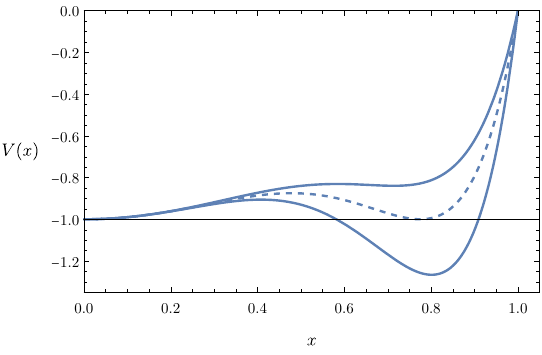}
\caption{
Plots of the potential $V(x)$ in eq.~\eqref{eq:universal_scaling_potential} for $\gamma |\q^2| = 0.02$ (top curve), $\gamma |\q^2| = 0.04$ (bottom curve), and $\gamma |\q^2| = 25\sqrt{15}/3456$ (dashed curve in the middle). In the limit \eqref{eq:c-limit}, potentials with $\gamma |\q^2| > 25\sqrt{15}/3456$ support bound states with energies $E<-1$, corresponding to causality-violating quasinormal modes with ${\rm Im}(\w) > |{\rm Im}(\q)|$. 
}
\label{fig:V}
\end{figure}

In the limit of small $\gamma$, the potentials $U_a^{(-2)}$ start at $O(\gamma)$; the expressions in Appendix~\ref{ap:us} show that these $O(\gamma)$ contributions are helicity-independent. The potentials $U_a^{(0)}$ start at $O(1)$ at small~$\gamma$; again, the expressions in Appendix~\ref{ap:us} show that these $O(1)$ contributions are helicity-independent. This helicity-independence at leading order gives a common bound-state construction for $a=0,1,2$. 
Consider the joint limit $\gamma\to0$ and $\hbar\to0$, with $\gamma/\hbar^2$ fixed and $E$ finite. Since $r_0=1/2$, eq.~\eqref{tknp} gives $4\pi^2T^2=1+\calo(\gamma)$. Consequently, for $0<x<1$, in the limit \eqref{eq:c-limit}, we have 
\begin{align}
\frac{U_a^{(-2)}}{\hbar^2}
&=-\frac{192\gamma}{\hbar^2}x^5(1-x^2)+\calo(\hbar^2),\notag\\
U_a^{(0)}&=-(1-x^2)+\calo(\hbar^2),
\qquad \hbar^2U_a^{(2)}=\calo(\hbar^2).
\label{eq:universal_scaling_terms}
\end{align}
The common leading-order potential in the limit \eqref{eq:c-limit} is therefore
\begin{align}
V(x)
&=-(1-x^2)\left(1+\frac{192\gamma}{\hbar^2}x^5\right),
\label{eq:universal_scaling_potential}\\
V(x)+1
&=x^2\left[1-\frac{192\gamma}{\hbar^2}x^3(1-x^2)\right].
\label{eq:universal_well_depth}
\end{align}
The kinetic energy operator in eq.~\eqref{eq:Scheq} is nonnegative, so the potential has to dip below $-1$ for a bound state with $E<-1$. Using
\begin{align}
\max_{0<x<1}x^3(1-x^2)=\frac{6\sqrt{15}}{125},
\qquad x_*^2=\frac35,
\end{align}
we find that the potential \eqref{eq:universal_scaling_potential} dips below $-1$ precisely when
\begin{align}
\gamma>\gamma_c^{(0)}(\hbar)
\equiv\frac{25\sqrt{15}}{3456}\,\hbar^2
=\frac{25\sqrt{15}}{3456\,|\q|^2}.
\label{eq:universal_well_threshold}
\end{align}
This is illustrated in Fig.~\ref{fig:V}. Again, here $\q$ is purely imaginary, with $|\q|=1/\hbar$. 
Standard quantum-mechanical arguments imply that in the limit $\hbar\to0$, the ground state energy will be at the minimum of the potential. Consequently, as the potential $V(x)$ stays finite in the limit \eqref{eq:c-limit}, there will be a bound state in that limit with $E<-1$, corresponding to causality-violating quasinormal modes with ${\rm Im}(\w) > |{\rm Im}(\q)|$.

The potential~\eqref{eq:universal_scaling_potential} can also be used to find the leading correction to $\gamma_c$ in the limit \eqref{eq:c-limit}. At the minimum $y_* = y(x_*)$ with $V=-1$,
\begin{align}
\left.\frac{d^2 V}{dy^2}\right|_{y_*}
=\frac{12}{5}\sqrt{\frac35},
\qquad x_*^5(1-x_*^2)=\frac{18\sqrt{15}}{625}.
\end{align}
Expanding $V(y)$ about $y_*$, the ground state energy shifts to $E = -1 + \hbar (\frac12 V''(y_*))^{1/2}$, due to the zero-point energy of the harmonic oscillator. The increased energy has to be compensated by increasing $\gamma/\hbar^2$ to lower the potential, which gives
\begin{align}
\gamma_c(\hbar)
&=\frac{25\sqrt{15}}{3456}\,\hbar^2
\left[1+\frac{\sqrt{6\sqrt{15}}}{3}\,\hbar
+O(\hbar^2)\right],
\label{eq:universal_quantum_threshold}\\
\gamma_c|\q|^2
&=\frac{25\sqrt{15}}{3456}
\left[1+\frac{\sqrt{6\sqrt{15}}}{3|\q|}
+O(|\q|^{-2})\right],\notag\\
|\q|_{\mathrm{onset}}&\sim
\sqrt{\frac{25\sqrt{15}}{3456\gamma}}.
\label{eq:universal_short_wavelength_onset}
\end{align}
Again, the violation of the condition ${\rm Im} (\w) \leq |{\rm Im}(\q)|$ which must be obeyed by quasinormal modes is a large-$|\q|$ effect: the onset of causality violation recedes to $|\q| \to\infty$ as $\gamma\to0$, rather than persisting at fixed momentum.

The above argument about the existence of states with $E<-1$ relies only on the leading $O(\gamma)$ term of $U_a^{(-2)}$ and the $O(\gamma^0)$ term of $U_a^{(0)}$. The argument is therefore insensitive to changes in the subleading terms in the potential that remain suppressed in the limit \eqref{eq:c-limit}. For example, an $O(\gamma^2)$ correction to $U_a^{(-2)}$ contributes only $O(\gamma^2/\hbar^2)=O(\hbar^2)$ to the potential. Extending this robustness to omitted higher-derivative terms in the ten-dimensional action~\eqref{s10} requires control of their momentum dependence. A hypothetical contribution to the potential \eqref{eq:uan} of the form
$\Delta U_a = U_a^{(-4)}/\hbar^4$, with $U_a^{(-4)}=O(\gamma^2)$
can survive when $\gamma/\hbar^2$ is fixed and can therefore modify the leading potential~\eqref{eq:universal_scaling_potential}. In section~\ref{sec:gamma2_extension} we test precisely this possibility for one explicit order-$\gamma^2$ extension of the bulk effective action in the helicity-2 channel. In that model the surviving contribution is attractive, and the bound state with $E<-1$ persists. This provides a nontrivial test of the instability, but it does not demonstrate that the instability persists in the full ten-dimensional effective action when all orders in the small $\alpha'$ expansion are taken into account.

\subsection{Unstable mode in the helicity-2 channel}
\label{sec:scalar_seed}

The quasinormal mode which corresponds to the ground state in the above quantum-mechanical problem can be constructed numerically, by solving the auxiliary one-dimensional Schr\"odinger equation~\eqref{eq:Scheq} at small but finite $\gamma$, without assuming the scaling limit~\eqref{eq:c-limit}. For the mode~\eqref{eq:wq-2}, the relation $\w(\q)$ may be viewed as the relation $\Gamma(v)$, which determines the instability rate of the spatially homogeneous mode in a moving reference frame as a function of the boost velocity. In order to find $\Gamma(v)$, one can proceed as follows. For a given $\gamma$, one fixes $v$, which determines the energy $E=-1/v^2$ by~\eqref{eq:hbarE}. Next, one tunes $\hbar$ until the normalizable near-boundary solution matches the decaying near-horizon solution. Once $\hbar$ is found, it determines $\Gamma$ via eq.~\eqref{eq:hbarE}.

The frequency, momentum, and growth rate are related to the tuned
value of $\hbar$ by
\begin{align}
\w=\frac{i}{\hbar v},
\qquad
\q=-\frac{i}{\hbar},
\qquad
\Gamma=\frac{\sqrt{1-v^2}}{\hbar v}\,,
\label{eq:scalar_seed_dictionary}
\end{align}
where we have assumed $v>0$.
To impose the boundary conditions in the quantum-mechanical problem,
we write
\begin{align}
\psi_2=x^{5/4}(1-x)^{1/(2\hbar v)}p(x),
\label{eq:scalar_seed_factorization}
\end{align}
where $p(x)$ approaches finite, nonzero constants at both endpoints.
We integrate from the two endpoints to an interior point and tune
$\hbar$ until the solutions and their first derivatives match, up to
an overall normalization. We select the solution with no radial
nodes, corresponding to the ground state.

\begin{figure}[t]
\centering
\includegraphics[width=\textwidth]{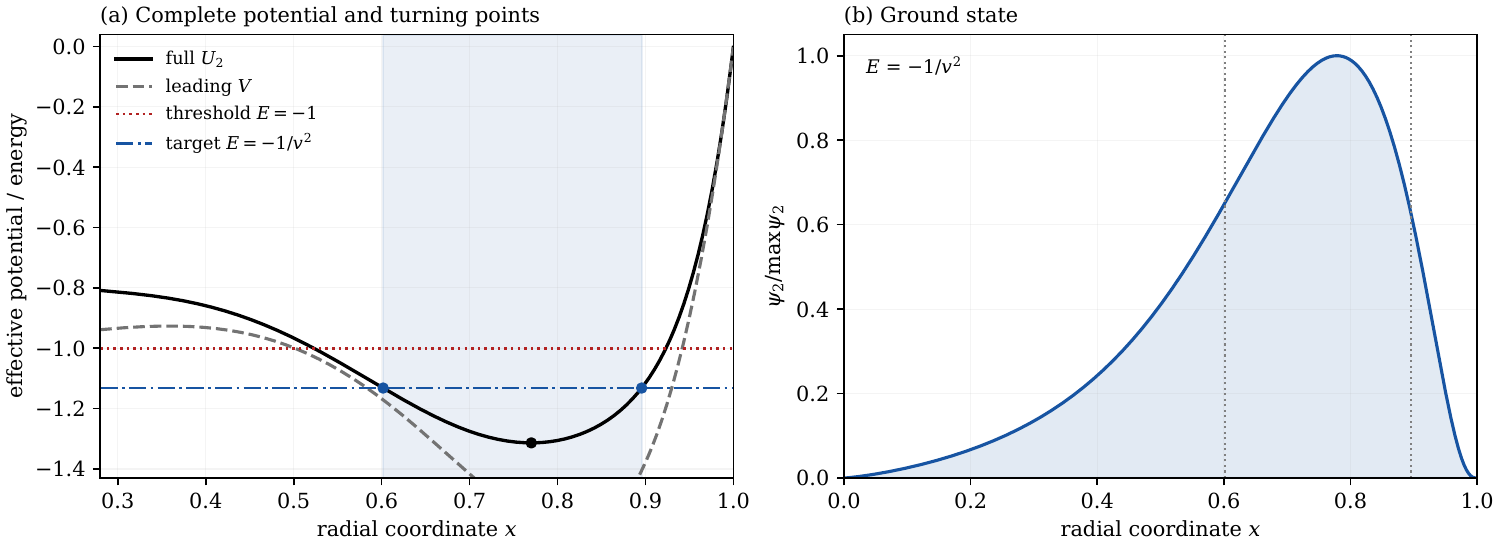}
\caption{
Helicity-2 bound state obtained by tuning $\hbar$ at the parameters
\eqref{eq:scalar_seed_parameters}--\eqref{eq:scalar_seed_hbar}.
Left: the full potential $U_2$ \eqref{eq:uan} (solid black), its leading scaling
approximation $V$ \eqref{eq:universal_scaling_potential} (dashed gray), the instability threshold $E=-1$
(dotted red), and the prescribed energy $E=-1/v^2$
(dash-dotted blue). The shaded interval is classically allowed.
Right: the ground-state wavefunction, normalized to unit maximum.
}
\label{fig:scalar_seed}
\end{figure}

\begin{figure}[t]
\centering
\includegraphics[width=\textwidth]{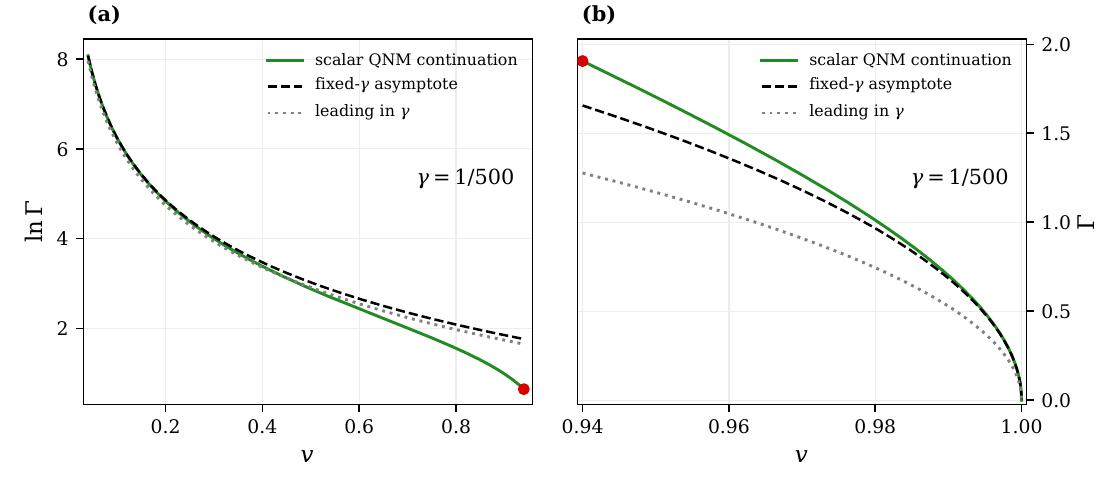}
\caption{
Growth rate of the helicity-2 instability at $\gamma=1/500$, corresponding to the 't~Hooft coupling in ${\cal N}=4$ SYM theory $\lambda\approx18$.
(a) The green curve shows $\ln\Gamma(v)$ as the bound state found from the full ``quantum-mechanical'' problem \eqref{eq:Scheq} is continued toward $v=0$. (b) The green curve shows $\Gamma(v)$ from the direct QNM calculation as the same branch is continued toward $v=1$. The red dots in both panels mark the bound state at $v=47/50$ displayed in Fig.~\ref{fig:scalar_seed}. The dashed black curves show the leading asymptotes at fixed $\gamma$, while the dotted gray curves retain only their leading small-$\gamma$ coefficients, as described in the text and derived in Appendices~\ref{app:scalar_small_v} and~\ref{app:scalar_v_one}.
}
\label{fig:scalar_continuation}
\end{figure}

Figure~\ref{fig:scalar_seed} illustrates this tuning. For
\begin{align}
\gamma=\frac{1}{500},
\qquad
v=\frac{47}{50},
\qquad
E=-\frac{2500}{2209},
\label{eq:scalar_seed_parameters}
\end{align}
the required value is
\begin{align}
\hbar\approx 0.190(4).
\label{eq:scalar_seed_hbar}
\end{align}
The full potential then admits the bound state at the prescribed
energy $E=-1/v^2$, and eq.~\eqref{eq:scalar_seed_dictionary} gives
$\Gamma\simeq1.906(0)$.
Next, keeping $\gamma=1/500$ fixed, we extend this computation to smaller $v$, repeating the tuning of $\hbar$ at each value of velocity and using the solution at the previous value of $v$ as the initial estimate. The resulting values of $\Gamma(v)$ form the green curve in Fig.~\ref{fig:scalar_continuation}(a). The red dot at $v=47/50$ marks the bound state shown in Fig.~\ref{fig:scalar_seed}. The solution remains nodeless along the green curve.

The unstable QNM becomes infinitely unstable as $v\to0$, with $\hbar = O(v)$, as we show in Appendix~\ref{app:scalar_small_v}. The small-$v$ behaviour is
\begin{align}
\Gamma &=\frac{D_2(\gamma)}{v^2}+\calo(v^0)\,,
\notag\\
D_2(\gamma)&=
\left[-\min_{0<x<1}
\left(
-\frac{192\gamma x^5(1-x^2)
[1-1200\gamma x^4(1-x^2)]}
{(1-30\gamma)
[1+6\gamma x^2(89x^4-119x^2+25)]}
\right)\right]^{-1/2}\,.
\label{eq:dashedblack}
\end{align}
This small-$v$ asymptotics is indicated by the black dashed curve in Fig.~\ref{fig:scalar_continuation}(a).
Retaining only the leading small-$\gamma$ term in the above $D_2$, one finds
\begin{align}
\Gamma &=\frac{C_0}{\sqrt{\gamma}\,v^2}\,,
\qquad
D_2(\gamma)=\frac{C_0}{\sqrt{\gamma}}
\left[1+\frac{14130}{343}\gamma+\calo(\gamma^2)\right]\,,
\notag\\
C_0&=\left[192\,\frac{2}{7}
\left(\frac{5}{7}\right)^{5/2}\right]^{-1/2}\,.
\label{eq:grey}
\end{align}
This small-$v$ approximation is indicated by the grey dotted curve in Fig.~\ref{fig:scalar_continuation}(a).
Note that taking $v\to0$ at fixed $\gamma$ takes $\gamma|\q|^2\to\infty$, outside the correlated limit~\eqref{eq:c-limit}. Consequently, the above small-$v$ asymptotics need not survive the inclusion of higher-derivative corrections omitted from our equation \eqref{eq:QNM_master_eq}.

Next, starting with the red dot in Fig.~\ref{fig:scalar_continuation}(a), which is given by eqs.~\eqref{eq:scalar_seed_parameters}, \eqref{eq:scalar_seed_hbar}, we continue the same QNM branch by increasing $v$ toward $v=1$, by solving the QNM equation \eqref{eq:QNM_master_eq} numerically. The result is shown in Fig.~\ref{fig:scalar_continuation}(b). The instability rate vanishes as $v\to1$, while $\operatorname{Im}\w$ and $|\q|$ approach a common finite value $\kappa_c$. As we show in Appendix~\ref{app:scalar_v_one}, the $v\to1$ behaviour is:
\begin{align}
\Gamma = \kappa_c(\gamma)\sqrt{1-v^2} \left( 1 + O(1{-}v^2) \right).
\label{eq:scalar_v_one_black}
\end{align}
We do not have a simple analytic formula $\kappa_c(\gamma)$, though it can be easily evaluated numerically. At $\gamma=1/500$, one has $\kappa_c(1/500)\approx 4.85450343$; the $v\to1$ asymptotics is indicated by the dashed black curve in Fig.~\ref{fig:scalar_continuation}(b).
The leading small-$\gamma$ contribution to $\kappa_c(\gamma)$,
\begin{align}
\Gamma =
\sqrt{\frac{25\sqrt{15}}{3456\gamma}}\,\sqrt{1-v^2},
\label{eq:scalar_v_one_gray}
\end{align}
is indicated by the dotted gray curve  in Fig.~\ref{fig:scalar_continuation}(b). 

\subsection{Connecting the unstable branch to Einstein quasinormal modes}
\label{sec:scalar_complex_momentum}

\begin{figure}[t]
\centering
\includegraphics[width=\textwidth]{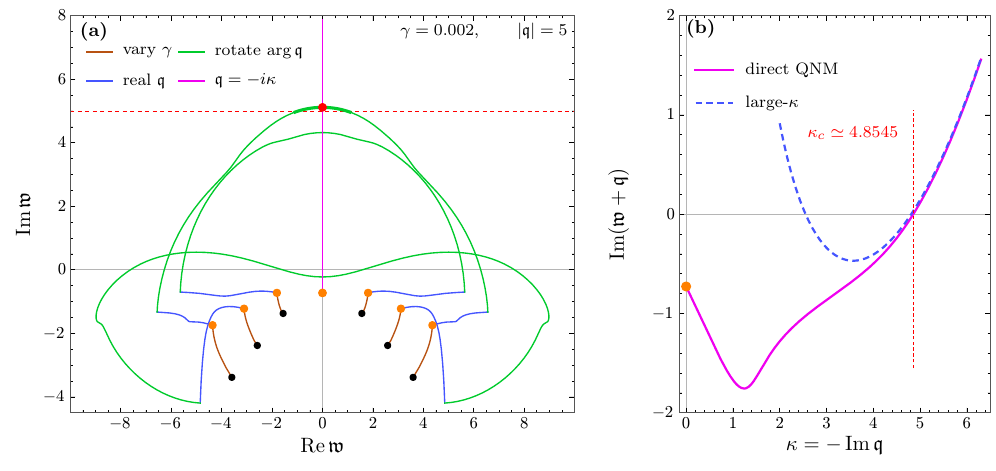}
\caption{Analytic continuation of scalar-channel QNMs to $\gamma=1/500$, corresponding to the 't~Hooft coupling in ${\cal N}=4$ SYM theory $\lambda\approx18$.
(a) Black dots are the modes at $\q=0$ obtained in Einstein gravity ($\gamma=0$). Brown curves show the trajectories of QNMs at $\q=0$, as $\gamma$ increases from $\gamma=0$ to $\gamma=1/500$, ending at orange dots. Blue curves show the trajectories of QNMs at $\gamma=1/500$, as real $\q$ increases from $\q=0$ to $\q=5$. Green curves show the trajectories of QNMs at $\gamma=1/500$ and $|\q|=5$, as the phase of $\q$ varies, while $|\q|=5$ stays fixed. The thick green segment at the top indicates the part of the QNM trajectory where the condition $\operatorname{Im}\w\leq|\operatorname{Im}\q|$ is violated. The magenta curve continues the central imaginary-frequency mode along $\q=-i\kappa$, with $\kappa>0$. Its red point at $\kappa=5$ is shared with the third-overtone ``Christmas-tree'' modes after a phase rotation. The horizontal dashed line marks $\operatorname{Im}\w=5$.
(b) The imaginary-momentum branch plotted as $\Delta(\kappa)=\operatorname{Im}(\w+\q)$: the magenta curve is the numerical QNM result and the dashed blue curve is the large-$\kappa$ approximation \eqref{eq:scalar_large_kappa_frequency}. The vertical dashed line marks $\kappa_c\simeq4.8545$; $\Delta>0$ signals instability in a boosted reference frame.}
\label{fig:scalar_complex_momentum}
\end{figure}

The imaginary-momentum instability can be connected to familiar Einstein-gravity QNMs by taking $\q$ complex, following the construction in~\cite{Buchel:2026rep}. At $\gamma=0$ and $\q=0$,
the first three scalar frequencies with negative real part are
\cite{Nunez:2003eq}
\begin{align}
\w_1&\simeq-1.559726-1.373338i,\notag\\
\w_2&\simeq-2.584760-2.381785i,\notag\\
\w_3&\simeq-3.593965-3.384782i.
\label{eq:scalar_einstein_modes}
\end{align}
In Fig.~\ref{fig:scalar_complex_momentum}(a), we follow these modes
through three consecutive steps: increase $\gamma$ from zero to
$1/500$ at $\q=0$;
increase real $\q$ from zero to $5$ at fixed $\gamma$; and
rotate the momentum through
\begin{align}
\q=5e^{-i\theta},\qquad 0\leq\theta\leq\frac{\pi}{2}.
\label{eq:scalar_momentum_rotation}
\end{align}
The reflected trajectories in Fig.~\ref{fig:scalar_complex_momentum}(a)
are obtained using the symmetry
$(\q,\w)\mapsto(-\q^*,-\w^*)$.

At $\gamma=1/500$, a further purely imaginary mode starts at
$\w\simeq-0.727717i$ when $\q=0$.\footnote{Such modes were identified and discussed in
\cite{Grozdanov:2016vgg}.} Continuing this mode along
$\q=-i\kappa$, with $\kappa\geq0$, reaches
\begin{align}
\w(\kappa=5)\simeq5.114658i.
\end{align}
This is also the endpoint of the $n=3$ rotation
\eqref{eq:scalar_momentum_rotation}; both continuations give
the same boundary-normalized radial solution. They therefore
connect the finite-coupling imaginary-frequency mode to an
Einstein QNM through the complex-momentum plane.
The $n=1$ and $n=2$ paths instead end at the distinct modes
$\w\simeq4.320290i$ and $\w\simeq-0.225142i$, respectively.
Although the green trajectories of the $n=1$ and $n=3$ modes
approach one another in the complex-frequency plane, the nearby
points correspond to different complex momenta and hence do not indicate
a degeneracy. 

For the branch with purely imaginary $\q$, shown by the magenta line in Fig.~\ref{fig:scalar_complex_momentum}(a), define
\begin{align}
\Delta(\kappa)\equiv\operatorname{Im}(\w+\q)
=\operatorname{Im}\w-\kappa.
\end{align}
As shown in figure~\ref{fig:scalar_complex_momentum}(b), this
quantity changes sign at $\kappa_c\simeq4.8545$.
For $\kappa>\kappa_c$, the mode obeys the instability criterion
and is precisely the branch followed in the preceding subsection:
\begin{align}
v=\frac{\kappa}{\operatorname{Im}\w}<1,
\qquad
\Gamma=\sqrt{(\operatorname{Im}\w)^2-\kappa^2}>0.
\end{align}
The large-$\kappa$ behavior is consistent with the small-$v$
asymptotics derived in Appendix~\ref{app:scalar_small_v}.
Including the first two subleading terms at $\gamma=1/500$ gives
\begin{align}
\operatorname{Im}\w
=\frac{\kappa^2}{D_2(1/500)}
-0.069042+\frac{8.83031}{\kappa^2}
+\calo(\kappa^{-4}),
\label{eq:scalar_large_kappa_frequency}
\end{align}
which, after subtracting $\kappa$, gives the dashed blue curve in Fig.~\ref{fig:scalar_complex_momentum}(b).
This approximation captures the large-momentum growth in panel (b); the crossing itself is determined from the full QNM equation. Thus the unstable branch is continuously connected (by changing $\gamma$ and $\q$), within the resummed order-$\gamma$ QNM equation \eqref{eq:QNM_master_eq}, to the ordinary Einstein-gravity QNM spectrum, while its small-$v$ end escapes to infinite frequency and momentum as described above.
Note that the causality violation exhibited by the ``non-perturbative'' mode on the imaginary axis at large $|\q|$ is qualitatively different from  the causality violation in the similar setup in Gauss-Bonnet gravity~\cite{Buchel:2026rep}; in the latter case, the ``non-perturbative'' frequency scales linearly with $|\q|$, while in our case it scales quadratically.

\section{Extension of the order-$\gamma$ model}
\label{sec:gamma2_extension}

\subsection{A phenomenological order-$\gamma^2$ extension}

The instability discussed above is obtained by treating the truncated order-$\gamma$ fluctuation equations at finite $\gamma$. To test its sensitivity to higher-order terms, we consider the phenomenological five-dimensional extension introduced in \cite{Buchel:2018eax},
\begin{align}
\widetilde S_5
=\frac{1}{2\kappa_5^2}\int d^5x\,\sqrt{-g}\,
\left[
R+\frac{12}{L^2}
+\gamma W+\alpha L^2\gamma^2W^2
\right].
\label{eq:gamma2_action}
\end{align}
Here $\alpha$ is a fixed dimensionless parameter of order one, with either sign. Contributions at order $\gamma^2$ arise both from evaluating the $\gamma W$ interaction on the order-$\gamma$ corrected background and fluctuations, and from the explicit $\alpha\gamma^2W^2$ interaction. The former contributions remain present even when $\alpha=0$.

This model is only a test of truncation sensitivity. It is not the complete ten-dimensional string effective action through order $\gamma^2$; in particular, it omits the order-$\gamma^{5/3}$ corrections mentioned after eq.~\eqref{s10}. We analyze only the helicity-2 channel. We expect analogous considerations to extend to the other helicity sectors, but have not verified this explicitly.

The background temperature, computed through order $\gamma^2$, obeys
\begin{align}
T_{\rm pert}^2
=\frac{r_0^2}{\pi^2}
\left[1+30\gamma+\frac{45}{7}(6699+2120\alpha)\gamma^2
+\calo(\gamma^3)\right].
\label{eq:gamma2_temperature_pert}
\end{align}
For the resummed fluctuation equation below, we use
\begin{align}
T&=\frac{r_0}{\pi}
\left[1-30\gamma-\frac{45}{7}(6559+2120\alpha)\gamma^2\right]^{-1/2},
\label{eq:gamma2_temperature}\\
\w&=\frac{\omega}{2\pi T},\qquad
\q=\frac{q}{2\pi T}.
\label{eq:gamma2_frequency_momentum}
\end{align}
This temperature agrees with \eqref{eq:gamma2_temperature_pert} through order $\gamma^2$ and makes the horizon indices of the resummed equation exactly $\pm i\w/2$. As in the order-$\gamma$ model, its unexpanded form specifies the finite-$\gamma$ normalization; it does not provide controlled thermodynamic information beyond the retained order.

The helicity-2 fluctuation equation can be reduced perturbatively to second order in derivatives through $O(\gamma^2)$:
\begin{align}
0=Z_2''+
\widetilde{\cal A}_2(x,\w,\q^2,\gamma)\,Z_2'
+\widetilde{\cal B}_2(x,\w,\q^2,\gamma)\,Z_2
+\calo(\gamma^3).
\label{eq:gamma2_reduced}
\end{align}
The coefficients $\widetilde{\cal A}_2$ and $\widetilde{\cal B}_2$ retain all terms through order $\gamma^2$. To obtain them, derivatives higher than second order in the order-$\gamma$ terms must be eliminated using the fluctuation equation through order $\gamma$; in the order-$\gamma^2$ terms, the Einstein equation suffices. Substituting the order-$\gamma$ correction back into the order-$\gamma$ terms generates additional order-$\gamma^2$ contributions. These are the iteration terms required for a consistent reduction. Explicit coefficients are given in Appendix~\ref{app:gamma2_coeffs}.

For the chosen fluctuation variable and radial coordinate, this procedure fixes the reduced coefficients uniquely through order $\gamma^2$. Equations differing by order-$\gamma^3$ terms are perturbatively equivalent to this accuracy, but can give different results when solved at finite $\gamma$. We use the coefficients in Appendix~\ref{app:gamma2_coeffs} without adding such terms and treat the resulting second-order equation nonperturbatively, as in the order-$\gamma$ analysis.

Using the radial coordinate $y$ defined in eq.~\eqref{eq:yx},
we write $Z_2=\psi_2/\widetilde{\cal C}_2$, where
\begin{align}
\frac{1}{\widetilde{\cal C}_2}
\frac{d\widetilde{\cal C}_2}{dy}
=\frac12\left(\frac{g'}{g^2}
+\frac{\widetilde{\cal A}_2}{g}\right).
\end{align}
Here and below primes denote $x$ derivatives. For purely imaginary frequency and momentum, defining
\begin{align}
\hbar^2=-\frac{1}{\q^2}>0,
\qquad
E=-\frac{\w^2}{\q^2}<0
\end{align}
gives
\begin{align}
-\hbar^2\frac{d^2\psi_2}{dy^2}
+\widetilde U_2\,\psi_2=E\psi_2,
\label{eq:gamma2_schrodinger}\\
\widetilde U_2
=E+\frac{\hbar^2}{g^2}
\left[
\frac{\widetilde{\cal A}_2'}{2}
+\frac{\widetilde{\cal A}_2^2}{4}
-\widetilde{\cal B}_2
+\frac{g''}{2g}-\frac{3(g')^2}{4g^2}
\right].
\label{eq:gamma2_liouville_potential}
\end{align}
No further expansion in $\gamma$ is made in this change of variables. In particular, squaring $\widetilde{\cal A}_2$ generates terms of order $\gamma^3$ and $\gamma^4$, even though the coefficients of the original second-order equation were truncated at order $\gamma^2$. These terms are fixed by the change of variables and do not represent additional interactions in the effective action.

The potential has the exact decomposition
\begin{align}
\widetilde U_2
={}&
\frac{1}{\hbar^6}\widetilde U_2^{(-6)}
+\frac{1}{\hbar^4}\widetilde U_2^{(-4)}
+\frac{1}{\hbar^2}\widetilde U_2^{(-2)}
+\widetilde U_2^{(0)}
+\hbar^2\widetilde U_2^{(2)}.
\label{eq:gamma2_potential_decomposition}
\end{align}
The leading small-$\gamma$ expressions for its components are given in Appendix~\ref{app:gamma2_potential}. The components $\widetilde U_2^{(-2)}$, $\widetilde U_2^{(0)}$ and $\widetilde U_2^{(2)}$ agree through order $\gamma$ with the ten-dimensional helicity-2 expressions in Appendix~\ref{ap:us}, including their explicit temperature dependence. This agreement applies only to the helicity-2 channel and the shared order $\gamma$; it does not establish an order-$\gamma^2$ equivalence with the ten-dimensional theory.

At fixed $\hbar$, $\widetilde U_2^{(-4)}$ starts at order $\gamma^2$, while $\widetilde U_2^{(-6)}$ starts at order $\gamma^4$. The former comes from the order-$\gamma^2\q^6$ term in $\widetilde{\cal B}_2$. The latter comes from squaring the order-$\gamma^2\q^4$ term in $\widetilde{\cal A}_2$ in eq.~\eqref{eq:gamma2_liouville_potential}. Thus its appearance is consistent with the order-$\gamma^2$ truncation of eq.~\eqref{eq:gamma2_reduced}.

\subsection{Scaling limit and persistence of the bound state}
\label{sec:gamma2_scaling}

We now repeat the correlated small-coupling limit \eqref{eq:c-limit},
\begin{align}
\gamma\to0,\qquad
\hbar\to0,\qquad
\sigma\equiv\frac{\gamma}{\hbar^2}
=\gamma|\q|^2
\quad\text{fixed},
\label{eq:gamma2_scaling}
\end{align}
with $E$ and $\alpha$ fixed. Here $\q$ is normalized with the temperature of the extended model, eq.~\eqref{eq:gamma2_temperature}. For $0<x<1$, in the region containing the potential well, we obtain
\begin{align}
  \frac{\widetilde U_2^{(-6)}}{\hbar^6}
  &=\calo(\gamma),
  \notag\\
  \frac{\widetilde U_2^{(-4)}}{\hbar^4}
  &=-36864\sigma^2 x^{10}(1-x^2)+\calo(\gamma),
  \notag\\
  \frac{\widetilde U_2^{(-2)}}{\hbar^2}
  &=-192\sigma x^5(1-x^2)+\calo(\gamma),
  \notag\\
  \widetilde U_2^{(0)}
  &=-(1-x^2)+\calo(\gamma),
  \qquad
  \hbar^2\widetilde U_2^{(2)}=\calo(\gamma).
\label{eq:gamma2_scaling_terms}
\end{align}
In particular, $\widetilde U_2^{(-6)}$ itself remains of order $\gamma^4$; its contribution to the potential is of order $\gamma^4/\hbar^6=\calo(\gamma)$ because $\hbar^2\sim\gamma$.

The order-one potential in this limit is therefore
\begin{align}
\boxed{
\widetilde V_2(x;\sigma)
=-(1-x^2)
\left[
1+192\sigma x^5+(192\sigma x^5)^2
\right].
}
\label{eq:gamma2_scaling_potential}
\end{align}
It is independent of both $E$ and $\alpha$. The full equation does depend on $\alpha$, as shown explicitly by the polynomials in Appendix~\ref{app:gamma2_coeffs}. However, their $\alpha$-dependent contributions to the potential vanish in the correlated limit for fixed $\alpha$. The additional order-one term,
\begin{align}
-192^2\sigma^2x^{10}(1-x^2),
\end{align}
comes from the order-$\gamma^2$ contributions generated by the $\gamma W$ interaction and remains present at $\alpha=0$. It does not come from the explicit $\alpha\gamma^2W^2$ interaction. Relative to \eqref{eq:universal_scaling_potential}, retaining the order-$\gamma^2$ terms therefore deepens the leading potential well.

The condition that this potential dip below $E=-1$ is
\begin{align}
\sigma>\sigma_c^{\rm ext},
\qquad
\sigma_c^{\rm ext}
=\min_{0<x<1}
\frac{
\sqrt{\dfrac{1+3x^2}{1-x^2}}-1
}{384x^5}.
\label{eq:gamma2_threshold_min}
\end{align}
Defining $z_c=192\sigma_c^{\rm ext}x_c^5$, the relevant positive root of
\begin{align}
2z_c^3+4z_c^2-6z_c-3=0
\label{eq:gamma2_zc}
\end{align}
determines
\begin{align}
x_c^2
=\frac{z_c(1+z_c)}{1+z_c+z_c^2},
\qquad
\sigma_c^{\rm ext}
=\frac{z_c}{192x_c^5}.
\label{eq:gamma2_sc_analytic}
\end{align}
Numerically,
\begin{align}
z_c\simeq1.275,\qquad
x_c\simeq0.862,\qquad
\sigma_c^{\rm ext}\simeq1.39\times10^{-2}.
\end{align}
For comparison, the order-$\gamma$ truncation gave
\begin{align}
\sigma_c^{(1)}
=\frac{25\sqrt{15}}{3456}
\simeq2.80\times10^{-2}.
\end{align}
Consequently the leading onset in the extended model is
\begin{align}
|\q|_{\rm onset}
=\sqrt{\frac{\sigma_c^{\rm ext}}{\gamma}}
\left[1+O(\sqrt{\gamma})\right],
\label{eq:gamma2_qonset}
\end{align}
to be compared with
$\sqrt{25\sqrt{15}/(3456\gamma)}
\approx 0.167/\sqrt{\gamma}$ in the order-$\gamma$ truncation.

For $\sigma>\sigma_c^{\rm ext}$, the leading potential has a
minimum below $-1$. At any fixed finite $\alpha$, the
$\alpha$-dependent corrections and the kinetic contribution
vanish as $\gamma\to0$. The bound state with $E<-1$ therefore
persists for sufficiently small $\gamma$, regardless of the
sign of $\alpha$.

Thus the $E<-1$ bound state persists in the order-$\gamma^2$ model \eqref{eq:gamma2_action} for arbitrary fixed $\alpha$. This conclusion is nevertheless truncation dependent. The scaling \eqref{eq:gamma2_scaling} enhances higher powers of momentum: for example, a term proportional to $\gamma^3/\hbar^6$ would remain of order one. The present calculation does not determine the sign or radial dependence of such terms in the complete string effective action. Higher-derivative corrections could therefore qualitatively modify, or even eliminate, the leading potential well. The result establishes persistence within the specified model, not in the complete finite-coupling theory.

\section{Discussion}
\label{sec:discussion}

Our results raise several related points.

\begin{itemize}
\item \textit{Resummation and the scope of the small-coupling argument.}
Solving the $O(\gamma)$-truncated fluctuation equation at finite $\gamma$ retains a particular set of higher-order effects, rather than the complete higher-derivative expansion of \eqref{s10}. The results obtained from such a ``resummation'' prescription beyond the strict perturbative $O(\gamma)$ order are therefore not universal (i.e., they depend on the implementation of the truncation), and are (unsurprisingly) sensitive to higher-derivative corrections~\cite{Waeber:2015oka,Buchel:2018eax}. The extension studied in section~\ref{sec:gamma2_extension} provides an explicit test of this sensitivity. In the helicity-2 channel, the higher-derivative correction to eq.~\eqref{eq:QNM_master_eq} generates an $O(\gamma^2)$ contribution $\hbar^{-4}\widetilde U_2^{(-4)}$ to the potential which survives in the correlated limit \eqref{eq:c-limit}. Thus this is precisely an example of a correction which cannot be discarded merely because $\gamma\to0$.

Within the particular extended model \eqref{eq:gamma2_action}, this new leading-order contribution to the potential is attractive. The resulting leading-order potential in the limit \eqref{eq:c-limit} is deeper than in the $O(\gamma)$ truncation, and the $E<-1$ bound state persists for every fixed finite $\alpha$ (in particular, for $\alpha=O(1)$ with either sign) at sufficiently small positive $\gamma$. This demonstrates that the instability is not removed by this particular $O(\gamma^2)$ deformation of the effective action.

It does not, however, establish robustness with respect to the full string effective action. Higher-order operators can acquire their own momentum enhancements in the same correlated limit; for example, contributions schematically proportional to $\gamma^3/\hbar^6$ need not vanish. We presently have no argument fixing the sign or functional form of the complete set of such terms, and hence no argument excluding the possibility that $\gamma^3$-like or still higher-derivative corrections could qualitatively alter or even eliminate the instability. The finite-$\gamma$ spectra discussed here should therefore be understood as properties of the specified truncated equations when they are viewed in the ``resummed'' sense.

\item \textit{Momentum dependence and the comparison with Gauss--Bonnet gravity.}
The term in the potential which is responsible for the distinctive momentum scaling of the instability we have discussed in this paper is the term $\hbar^{-2}U_a^{(-2)}$. Since $\hbar^{-2}=|\q|^2$, the attractive contribution to the potential which is responsible for $E$ dropping below $-1$ is proportional to $\gamma|\q|^2$, and the onset of causality violation obeys~\eqref{eq:universal_short_wavelength_onset}. Within the specified $O(\gamma)$-truncated equations, for every sufficiently small positive $\gamma$, causality violation (and the corresponding instability) can therefore be reached by increasing $|\q|$, with the onset receding as $|\q|\sim\gamma^{-1/2}$ when $\gamma\to0$.
In the Gauss-Bonnet model discussed in~\cite{Buchel:2026rep}, the contribution $U_a^{(-2)}$ is absent: the leading large-momentum potential
depends on the Gauss-Bonnet coupling, and the condition $E<-1$ sets a coupling threshold rather than a condition on $\gamma|\q|^2$. This difference with the Gauss-Bonnet theory pertains to the leading large-momentum instability criterion; finite-momentum corrections and instability growth rates in Gauss-Bonnet gravity still depend on momentum.

\item \textit{A thermal-scale growth rate at small $\gamma$.}
In a boosted reference frame, the instability need not be confined to large frequency, even when $\gamma$ is small. At representative values \eqref{eq:scalar_seed_parameters}, \eqref{eq:scalar_seed_hbar}, with $\gamma=1/500$ and $v=47/50$, the homogeneous boosted mode is
\begin{align}
(\w',\q')\approx (1.906(0)\,i,0).
\label{eq:discussion_boosted_seed}
\end{align}
Using the normalization \eqref{defwq}, its e-folding time is
\begin{align}
\tau_{\mathrm{grow}}=\frac{1}{2\pi T\Gamma}
\approx \frac{0.0835}{T}.
\end{align}
Thus a small coefficient of the higher-derivative correction already permits rapid unstable growth of a spatially homogeneous perturbation in an appropriate boosted frame. 
This example is consistent with the short-wavelength origin of the effect: the same mode has nonzero imaginary momentum in the fluid rest frame. There is no strictly positive minimum boost velocity in our leading small-$\gamma$ construction: every fixed $0<v<1$ admits the unstable branch at sufficiently small positive $\gamma$ and sufficiently large imaginary momentum. The scalar-channel small-$v$ asymptotics \eqref{eq:scalar_small_v_asymptote} make the limiting mechanism explicit: $|\q|\sim D_2(\gamma)/v$ and $\Gamma\sim D_2(\gamma)/v^2$, diverging as $v\to0$. Again, these statements concern the specified truncated equation, with solutions understood in the ``resummed'' sense; the small-$\gamma$ expansion is not uniform as $v\to0$, and does not establish the same conclusion for the full string theory spectrum.

\item \textit{The large-momentum limit and the string scale.}
There is a qualitative limitation to the ``non-perturbative'' approach to the perturbatively derived equations, arising from the fact that supergravity is an effective description, only valid below the string scale. Using $\gamma \sim \lambda^{-3/2}$, the correlated limit \eqref{eq:c-limit} with $\gamma|\q|^2=\text{fixed}$ probes
\begin{align}
  |\q|\sim \gamma^{-1/2} \sim \lambda^{3/4} \,.
\end{align}
As the energy $E = -\w^2/\q^2$ stays fixed in the limit \eqref{eq:c-limit}, one also has 
\begin{align}
  \frac{\omega}{T} \sim  \lambda^{3/4} \,.
\label{eq:ws}
\end{align}
On the other hand, a fixed massive string level has
\begin{align}
m_sL\sim\lambda^{1/4} \,,
\label{eq:ms}
\end{align}
in AdS units \cite{Aharony:1999ti}. A massive field with mass $m \gg 1/L$ in AdS gives rise to quasinormal modes with $\omega/T \sim mL$, see e.g.~\cite{Festuccia:2008zx}. While one can not directly interpret eq.~\eqref{eq:ms} as implying, by itself, the existence of a quasinormal pole at $\omega/T \sim \lambda^{1/4}$, it is clear that QNMs with frequencies scaling as \eqref{eq:ws} lie parametrically above the string scale. Thus,  causality violation which we have identified in the truncated theory \eqref{eq:QNM_master_eq} is physically unreliable because string excitations (ignored in the supergravity approximation) can not be neglected at the relevant scale. The ${\cal N}{=}4$ SYM theory does not become causality-violating at finite 't~Hooft coupling, another reason not to take at face value the ``non-perturbative'' results obtained in the bulk theory which is only perturbatively controlled. 

The above comments on the (un)reliability of the supergravity approximation also apply to the modes at $\q=0$. Ref.~\cite{Grozdanov:2016vgg} has argued for the scaling of the $\q=0$ mode on the imaginary axis of the form $\w\sim 1/\gamma$, when $\gamma$ is not too small. Such a scaling corresponds to $\omega/T \sim \lambda^{3/2}$, placing the ``non-perturbative'' mode parametrically even further beyond the validity regime of supergravity than the scale of causality violation exhibited by the same ``non-perturbative'' mode at small $\gamma$ and $|\q| \gtrsim 0.17/\sqrt{\gamma}$. 

\item
It is important to note that, in the ``resummed'' approach to black brane perturbations, the ``extra'', or ``non-perturbative'' mode on the imaginary axis for the black brane at rest is not the only pathological mode at non-zero~$\gamma$. As emphasized in \cite{Grozdanov:2019kge, Grozdanov:2019uhi}, different quasinormal modes form different branches of the same multi-valued complex function, a phenomenon which persists at non-zero~$\gamma$~\cite{Grozdanov:2021jfw}. Thus, the conventional ``Christmas-tree'' modes can become causality violating as well. For example, as discussed in sec.~\ref{sec:scalar_complex_momentum}, at $\gamma=1/500$, already the third ``Christmas-tree'' overtone for the black brane at rest becomes causality-violating. 

\item
The causality violation exhibited by the truncated theory \eqref{eq:QNM_master_eq} is, in a sense, significantly worse than the causality violation observed in Gauss-Bonnet gravity in ref.~\cite{Buchel:2026rep}. In the latter case, causality-violating modes for the black brane at rest had $\w$ proportional to $|\q|$ at large $|\q|$, while here we have $\w$ proportional to $|\q|^2$ at large $|\q|$, consistent with the extra $\q$-dependence which was not present in the analysis of~\cite{Buchel:2026rep}.

\end{itemize}

\paragraph{Acknowledgements.} 
This work was supported in part by NSERC of Canada through the Discovery Grants program.

\appendix

\section{QNM equation coefficients}\label{app:QNM_coeffs}
The dimensionless frequency and momentum are defined by \eqref{defwq}, where the temperature $T$ is given by the first equality in eq.~\eqref{tknp}.
In the coefficients ${\cal A}_a$, ${\cal B}_a$ below, we set $r_0=1/2$. For arbitrary $r_0$, replace $T$ by $T/(2r_0)$ in these expressions.
The gauge-invariant combinations $Z_a$ in the different helicity sectors of the metric fluctuations $\xi_{MN}$ are defined as follows~\cite{Kovtun:2005ev,Buchel:2004di,Benincasa:2005iv,Benincasa:2005qc}:%
\begin{subequations}
  \begin{align}
{\rm scalar\ channel:}\qquad Z_2 &= \xi_1{}^2\,,
\\
{\rm shear\ channel:}\qquad Z_1 &= \q\,\xi_0{}^1+\w\,\xi_3{}^1\,,
\\
  {\rm sound\ channel:}\qquad Z_0 &=
{4}\frac{\q}{\w}\,\xi_0{}^3+2\xi_3{}^3
-\left(\xi_1{}^1+\xi_2{}^2\right)
\left(
1-\frac{\q^2}{\w^2}
\frac{\hat c_1'\hat c_1}{\hat c_2'\hat c_2}
\right)
-2\frac{\q^2}{\w^2}
\frac{\hat c_1^2}{\hat c_2^2}\,\xi_0{}^0
  \end{align}
\end{subequations}
where $\xi\equiv \xi_1{}^1+\xi_2{}^2$, and spatial indices are raised with the
effective five-dimensional Einstein frame background geometry 
\begin{equation}
d{\hat s}_5^2=-{\hat c}_1^2\ dt^2+{\hat c}_2^2\ (dx^2+dy^2+dz^2)+{\hat c}_3^2\ dr^2\,,\qquad {\hat c}_i\equiv c_4^{5/3}\ c_i\,,
\label{5dm}
\end{equation}
$\xi_M{}^N\equiv {\hat g}^{L N}\xi_{M L}$.
The coefficients $c_i$ are given by eq.~\eqref{defm}.

\subsection{Helicity-2 (scalar channel)}
\label{app:QNM_coeffs_scalar}
The coefficients of the QNM equation~\eqref{eq:QNM_master_eq} in the scalar channel are given by
\begin{subequations}
\label{eq:scalar_AB}
\begin{align}
{\cal A}_2(x) &= -\frac{1+x^2}{x(1-x^2)}
+\gamma\left[3840\pi^2 T^2\q^2 x^4
+\frac{x}{4}\left(3171x^4+2306x^2-600\right)\right],
\\
{\cal B}_2(x) &= \frac{4\pi^2 T^2\left[\w^2-\q^2(1-x^2)\right]}
{x(1-x^2)^2}
\nonumber\\
&\quad +\gamma\Biggl[
\frac{24\pi^2 T^2\w^2 x\left(89x^4-119x^2+25\right)}{(1-x^2)^2}
\nonumber\\
&\qquad -\frac{4\pi^2 T^2\q^2 x\left(851x^4-789x^2+75\right)}{1-x^2}
+\frac{3072\pi^4 T^4\q^4 x^4}{1-x^2}
\nonumber\\
&\qquad -\frac{25x^2\left(9x^4-2x^2-2\right)}{4(1-x^2)}
\Biggr].
\end{align}
\end{subequations}

The logarithmic derivative of ${\cal C}_2(x)$ is given by
\begin{equation}
\label{eq:scalar_logC2}
\frac{d}{dx}\ln {\cal C}_2(x)
=\frac{(30\gamma-1)\left[\gamma x^2\left(3171x^4+2306x^2-600\right)-6\right]
-3840\gamma\q^2 x^5}{8x(30\gamma-1)}\,.
\end{equation}

\subsection{Helicity-1 (shear channel)}
\label{app:QNM_coeffs_shear}
For compactness, define
\begin{align}
\Delta_1(x) &\equiv \w^2-\q^2(1-x^2),
\\
P_1(x) &\equiv \w^4\left(3171x^4+2306x^2-600\right)
\nonumber\\
&\quad -2\q^2\w^2\left(1773x^6-5471x^4+4058x^2-300\right)
\nonumber\\
&\quad +5\q^4x^2(1-x^2)^2\left(1162-849x^2\right),
\\
Q_1(x) &\equiv 5\w^4+\q^2\w^2\left(31x^2-25\right)
+20\q^4(1-x^2)^2.
\end{align}
The coefficients of the QNM equation~\eqref{eq:QNM_master_eq} in the shear channel are given by
\begin{subequations}
\label{eq:shear_AB}
\begin{align}
{\cal A}_1(x) &= -\frac{1}{x}-\frac{2\w^2 x}{(1-x^2)\Delta_1}
+\frac{\gamma x}{4\Delta_1^2}
\left[P_1+1024\pi^2T^2\q^2x^3Q_1\right],
\\
{\cal B}_1(x) &= \frac{4\pi^2T^2\Delta_1}{x(1-x^2)^2}
+\gamma\Biggl[\frac{3072\pi^4T^4\q^4x^4}{1-x^2}
\nonumber\\
&\qquad +\frac{60\pi^2T^2\q^4x\left(85x^4-91x^2+5\right)}{\Delta_1}
\nonumber\\
&\qquad -\frac{36\pi^2T^2\q^2\w^2x\left(233x^4-231x^2+25\right)}{(1-x^2)\Delta_1}
\nonumber\\
&\qquad +\frac{24\pi^2T^2\w^4x\left(89x^4-119x^2+25\right)}{(1-x^2)^2\Delta_1}
\nonumber\\
&\qquad -\frac{25x^2\left[2\q^2(1-x^2)^2(3x^2+1)
+\w^2(9x^4-2x^2-2)\right]}{4(1-x^2)\Delta_1}
\Biggr],
\end{align}
\end{subequations}
where the arguments of $\Delta_1$, $P_1$, and $Q_1$ have been suppressed.

The logarithmic derivative of ${\cal C}_1(x)$ is given by
\begin{equation}
\label{eq:shear_logC2}
\frac{d}{dx}\ln {\cal C}_1(x)
=\frac{(30\gamma-1)\left[\gamma x^2P_1-2\Delta_1(3\Delta_1 +4\q^2x^2)\right]
-256\gamma\q^2x^5Q_1}{8x(30\gamma-1)\Delta_1^2}\,.
\end{equation}

\subsection{Helicity-0 (sound channel)}
\label{app:QNM_coeffs_sound}
For compactness, define
\begin{equation}
\Delta_0(x)\equiv \q^2x^2-3\q^2+3\w^2,
\end{equation}
and the polynomials
\begin{align}
P_0(x) &\equiv \Delta_0^3
\left(5120\pi^2T^2\q^2x^3+9513x^4+6918x^2-1800\right)
\nonumber\\
&\quad -16\q^2\Delta_0^2\Bigl[1280\pi^2T^2\q^2x^3(9-7x^2)
\nonumber\\
&\hspace{38mm} +3636x^6-4104x^4+1728x^2+225\Bigr]
\nonumber\\
&\quad -16\q^4x^2\Delta_0\Bigl[512\pi^2T^2\q^2x^3(5x^2-9)
\nonumber\\
&\hspace{38mm} +23571x^6-68904x^4+60480x^2-15777\Bigr]
\nonumber\\
&\quad +9216\q^6x^4(1-x^2)^2(17x^2-27),
\\
R_0(x) &\equiv 2048\pi^2T^2\q^2x(88x^4-189x^2+81)
\nonumber\\
&\quad +3\left(43341x^6-74255x^4+25994x^2+1200\right),
\\
S_0(x) &\equiv -256\pi^2T^2\q^2x(41x^4-114x^2+81)
\nonumber\\
&\quad +3\left(6225x^6-18744x^4+17568x^2-5259\right).
\end{align}
The coefficients of the QNM equation~\eqref{eq:QNM_master_eq} in the sound channel are given by
\begin{subequations}
\label{eq:sound_AB}
\begin{align}
{\cal A}_0(x) &= -\frac{1+x^2}{x(1-x^2)}
-\frac{4\q^2x}{\Delta_0}
+\frac{\gamma xP_0}{12\Delta_0^3},
\\
{\cal B}_0(x) &=
\frac{4\pi^2T^2\left[\w^2-\q^2(1-x^2)\right]}{x(1-x^2)^2}
-\frac{4\q^2x^2}{(1-x^2)\Delta_0}
\nonumber\\
&\quad +\gamma\Biggl\{
\frac{3072\pi^4T^4\q^4x^4}{1-x^2}
\nonumber\\
&\qquad +\frac{4\pi^2T^2x}{3(1-x^2)^2}
\Bigl[6\Delta_0(89x^4-119x^2+25)
\nonumber\\
&\hspace{43mm} +\q^2(4867x^6-7756x^4+2604x^2+225)\Bigr]
\nonumber\\
&\qquad -\frac{25x^2(9x^4-2x^2-2)}{4(1-x^2)}
-\frac{\q^2x^2R_0}{6(1-x^2)\Delta_0}
-\frac{4\q^4x^4S_0}{3(1-x^2)\Delta_0^2}
\nonumber\\
&\qquad +\frac{768\q^6x^6(1-x^2)(17x^2-27)}{\Delta_0^3}
\Biggr\},
\end{align}
\end{subequations}
where the arguments of $\Delta_0$, $P_0$, $R_0$, and $S_0$ have been suppressed.
The logarithmic derivative of ${\cal C}_0(x)$, with
$T=1/(2\pi\sqrt{1-30\gamma})$ substituted, is given by
\begin{equation}
\label{eq:logderiv_C3}
\begin{aligned}
\frac{d}{dx}\ln {\cal C}_0(x)
&=\frac{1}{24x(1-30\gamma)\Delta_0^3}
\Bigl\{(1-30\gamma)
\left[\gamma x^2\widehat P_0-6\Delta_0^2(3\Delta_0+8\q^2x^2)\right]
\\
&\qquad +256\gamma\q^2x^5\Delta_0
\left[5\Delta_0^2+20\q^2\Delta_0(7x^2-9)
+8\q^4x^2(9-5x^2)\right]\Bigr\},
\end{aligned}
\end{equation}
where the temperature-independent polynomial $\widehat P_0(x)$ is
\begin{equation}
\begin{aligned}
\widehat P_0(x)
&\equiv \Delta_0^3(9513x^4+6918x^2-1800)
\\
&\quad -16\q^2\Delta_0^2(3636x^6-4104x^4+1728x^2+225)
\\
&\quad -16\q^4x^2\Delta_0(23571x^6-68904x^4+60480x^2-15777)
\\
&\quad +9216\q^6x^4(1-x^2)^2(17x^2-27).
\end{aligned}
\end{equation}

\section{Leading-order in $\gamma$ potential terms $U_a^{(n)}$}\label{ap:us}
The dimensionless frequency and momentum are defined by \eqref{defwq}, where the temperature $T$ is given by the first equality in eq.~\eqref{tknp}. 
In the expressions below, note that to leading order in $\gamma$, $U_a^{(-2)}$ and $U_a^{(0)}$ are the same across all helicity channels, while $U_a^{(2)}$ differ. It is convenient to use the compact radial coordinate $x$; the relation between it and the $y$-radial coordinate is given by~\eqref{eq:yx}.

The expressions below display terms through linear order in the explicit $\gamma$ dependence, while keeping $T$ unexpanded as in eq.~\eqref{tknp}. The $O(\gamma^2)$ remainders denote known terms generated by the exact transformation of eq.~\eqref{eq:QNM_master_eq} to Schr\"odinger form, which are straightforward to obtain but too lengthy to write down. The full transformed potential, including these $O(\gamma^2)$ terms, is used in the finite-$\gamma$ computations, such as calculations in Appendix~\ref{app:scalar_asymptotics}.

\subsection{Helicity-2 (scalar channel)}
\label{app:us_scalar}

\begin{align}
U_2^{(-2)}
&=-3072\pi^4 T^4\gamma x^5(1-x^2)
+\mathcal{O}(\gamma^2)\,,\label{eq:scalar_Uminus2}\\
U_2^{(0)}
&=(1-4\pi^2 T^2)E-4\pi^2 T^2(1-x^2) \notag\\
&\quad-4\gamma\pi^2 T^2 x^2
\Bigl[
6(89x^4-119x^2+25)E
+1549x^6-2200x^4+576x^2+75
\Bigr]
+\mathcal{O}(\gamma^2)\,,\label{eq:scalar_Uzero}\\
U_2^{(2)}
&=\frac{3(1-x^2)(3x^2+5)}{16x}
-6\gamma x^3(1-x^2)(387x^4-70x^2-119)
+\mathcal{O}(\gamma^2)\,.
\label{uscalar}
\end{align}

\subsection{Helicity-1 (shear channel)}
\label{app:us_shear}

\begin{align}
U_1^{(-2)}
&=-3072\pi^4 T^4\gamma x^5(1-x^2)
+\mathcal{O}(\gamma^2)\,,\\
U_1^{(0)}
&=(1-4\pi^2 T^2)E-4\pi^2 T^2(1-x^2)
-\frac{4\gamma\pi^2 T^2 x^2}{(1+E-x^2)^3}
\Bigl[
6(89x^4-119x^2+25)E^4 \notag\\
&\qquad +(1-x^2)(2365x^4-3027x^2+525)E^3 \notag\\
&\qquad +(1-x^2)(781x^6+2960x^4-4032x^2+675)E^2 \notag\\
&\qquad +(1-x^2)^2(2105x^6-2384x^4-864x^2+375)E \notag\\
&\qquad -15(1-x^2)^4(43x^4-37x^2-5)
\Bigr]
+\mathcal{O}(\gamma^2)\,,\\
U_1^{(2)}
&=\frac{1-x^2}{16x(1+E-x^2)^2}
\Bigl[
3(3x^2+5)E^2-2(25x^4+6x^2-15)E
-7x^6+13x^4-21x^2+15
\Bigr] \notag\\
&\quad-\frac{6\gamma x^3(1-x^2)}{(1+E-x^2)^3}
\Bigl[
(387x^4-70x^2-119)E^3 \notag\\
&\qquad +(616x^6-1484x^4+1276x^2-403)E^2 \notag\\
&\qquad +2(1-x^2)(527x^6-1341x^4+1071x^2-262)E \notag\\
&\qquad +120(1-x^2)^4(3x^2-2)
\Bigr]
+\mathcal{O}(\gamma^2)\,.
\label{ushear}
\end{align}

\subsection{Helicity-0 (sound channel)}
\label{app:us_sound}

\begin{align}
U_0^{(-2)}
&=-3072\pi^4 T^4\gamma x^5(1-x^2)
+\mathcal{O}(\gamma^2)\,,\\
U_0^{(0)}
&=(1-4\pi^2 T^2)E-4\pi^2 T^2(1-x^2)
-\frac{4\gamma\pi^2 T^2 x^2}{(3+3E-x^2)^3}
\Bigl[
162(89x^4-119x^2+25)E^4 \notag\\
&\qquad -9(6203x^6-15740x^4+11292x^2-1575)E^3 \notag\\
&\qquad +3(7213x^8-69171x^6+121376x^4-65853x^2+6075)E^2 \notag\\
&\qquad -3(8619x^{10}-43358x^8+113255x^6-130843x^4+55782x^2-3375)E \notag\\
&\qquad -(1-x^2)(4691x^{10}-35552x^8+85929x^6-105531x^4+50112x^2-2025)
\Bigr]
+\mathcal{O}(\gamma^2)\,,\\
U_0^{(2)}
&=\frac{1-x^2}{16x(3+3E-x^2)^2}
\Bigl[
27(3x^2+5)E^2-6(73x^4-12x^2-45)E \notag\\
&\qquad +9x^6-295x^4-9x^2+135
\Bigr] \notag\\
&\quad-\frac{6\gamma x^3(1-x^2)}{(3+3E-x^2)^4}
\Bigl[
81(387x^4-70x^2-119)E^4 \notag\\
&\qquad +108(1390x^6-1624x^4+1038x^2-403)E^3 \notag\\
&\qquad -18(23459x^8-86708x^6+93852x^4-39048x^2+5869)E^2 \notag\\
&\qquad +12(9856x^{10}-88658x^8+225220x^6-218377x^4+83822x^2-9939)E \notag\\
&\qquad -9333x^{12}+129150x^{10}-633149x^8+1282444x^6 \notag\\
&\qquad -1137927x^4+420774x^2-47511
\Bigr]
+\mathcal{O}(\gamma^2)\,.
\label{usound}
\end{align}

\section{Asymptotics of the unstable helicity-2 mode}
\label{app:scalar_asymptotics}

Here we derive the two asymptotics of the unstable helicity-2 growth rate,
\begin{align*}
 {\Gamma(v)}& {\sim\frac{D_2(\gamma)}{v^2},\qquad v\to0,}\\
 {\Gamma(v)}& {\sim\kappa_c(\gamma)\sqrt{1-v^2},\qquad v\to1},
\end{align*}
where $D_2(\gamma)$ and $\kappa_c(\gamma)$ are specified below. 
In each case, we first take the velocity limit at fixed $\gamma$, and then expand its coefficient at small $\gamma$. The full fixed-$\gamma$ coefficients give the black dashed curves in Fig.~\ref{fig:scalar_continuation}; their leading small-$\gamma$ contributions give the gray dotted curves.

\subsection{The limit $v\to0$}
\label{app:scalar_small_v}

Using the coefficients ${\cal A}_2(x)$ and ${\cal B}_2(x)$ of the QNM equation \eqref{eq:QNM_master_eq} given in Appendix~\ref{app:QNM_coeffs_scalar}, one can use the change of variables described in Sec.~\ref{sec:perturbation_equations} to rewrite the equation for $Z_2(x)$ in the form of the Schr\"odinger-like equation~\eqref{eq:Scheq} for $\psi_2(y)$. The resulting scalar potential in eq.~\eqref{eq:Scheq} is no longer linear in $\gamma$ (due to the variable change), but it is linear in $E$. 
In what follows it is convenient to represent the resulting potential~\eqref{eq:uan} as
\begin{align}
U_2=V_2+E(1-R_2),
\qquad
R_2=1-\frac{\partial U_2}{\partial E},
\label{eq:scalar_energy_split}
\end{align}
Here $V_2=U_2|_{E=0}$ is the energy-independent part of the potential, and $R_2$ collects its energy dependence; both $V_2$ and $R_2$ are independent of $E$. Note that the potential $V_2$ is distinct from the leading scaling-limit potential $V$ in eq.~\eqref{eq:universal_scaling_potential}. Moving the energy-dependent term to the right-hand side rewrites the Schr\"odinger equation~\eqref{eq:Scheq} as an eigenvalue problem with an energy-independent differential operator and weight $R_2$:
\begin{align}
\left[-\hbar^2\frac{d^2}{dy^2}+V_2\right]\psi_2
=E R_2\psi_2.
\label{eq:scalar_seed_generalized_spectrum}
\end{align}
Here $y$ is the radial coordinate defined in eq.~\eqref{eq:yx}.
Using the coefficients in Appendix~\ref{app:QNM_coeffs_scalar}, together with the unexpanded temperature prescription in eq.~\eqref{tknp} gives
\begin{align}
R_2
&=\frac{1+6\gamma x^2(89x^4-119x^2+25)}{1-30\gamma},
\notag\\
  V_2 = \frac{1}{\hbar^2} U_2^{(-2)} + O(\hbar^0) 
&=- \frac{1}{\hbar^2} \frac{192\gamma x^5(1-x^2)
\left[1-1200\gamma x^4(1-x^2)\right]}{(1-30\gamma)^2} + O(\hbar^0).
\label{eq:scalar_small_v_coefficients}
\end{align}
These expressions retain the higher powers of $\gamma$, generated by changing variables in eq.~\eqref{eq:QNM_master_eq}, in particular by the terms quadratic in ${\cal A}_2$ which arise when eliminating the first derivative. No expansion in $\gamma$ is performed in going from $Z_2(x)$ to $\psi_2(y)$; eq.~\eqref{eq:QNM_master_eq}, with the coefficients given in Appendix~\ref{app:QNM_coeffs_scalar}, is viewed as an exact equation at small but non-zero values of $\gamma$ in the ``resummed'' approach to black brane perturbations. Note that $R_2>0$ throughout $0<x<1$ for sufficiently small positive $\gamma$ (including the value $\gamma=1/500$ used in Fig.~\ref{fig:scalar_continuation}).

Recall that $E=-1/v^2$. The limit $v\to0$ (corresponding to $E\to-\infty$) can be reached by $\hbar\to0$. Indeed, as $\hbar\to0$, the kinetic-energy contribution to the ground state energy is subleading relative to the $O(1/\hbar^2)$ potential term.%
\footnote{
By a variational argument, with a trial wave function whose width is of order $O(\hbar)$ in the $\hbar\to0$ limit.
}
In the $\hbar\to0$ limit, the ground-state energy is then given by the minimum of the corresponding ratio, 
\begin{align}
  -\frac{1}{v^2} = \frac{1}{\hbar^2} \left( \min_{0<x<1} F_{2}(x) \right) + O(1)\,,
\end{align}
where
\begin{align}
  F_{2}(x) \equiv\frac{U_2^{(-2)}(x)}{R_2(x)}
=-\frac{192\gamma x^5(1-x^2)
\left[1-1200\gamma x^4(1-x^2)\right]}
{(1-30\gamma)
\left[1+6\gamma x^2(89x^4-119x^2+25)\right]}.
\label{eq:scalar_small_v_potential}
\end{align}
The minimum can be taken over $x$ because $y(x)$ is monotonic.
Defining
\begin{align}
D_2(\gamma)\equiv
\left[-\min_{0<x<1}F_2\right]^{-1/2},
\label{eq:scalar_small_v_coefficient}
\end{align}
and using eq.~\eqref{eq:scalar_seed_dictionary}, we obtain
\begin{align}
\hbar\sim\frac{v}{D_2(\gamma)},
\qquad
|\q|\sim\frac{D_2(\gamma)}{v},
\qquad
\Gamma(v)=\frac{D_2(\gamma)}{v^2}
\left[1+\calo(v^2)\right].
\label{eq:scalar_small_v_asymptote}
\end{align}
Thus, the small-$\hbar$ limit is the small-$v$ limit; the above small-$v$ approximation to $\Gamma$ is shown by the dashed black curve in Fig.~\ref{fig:scalar_continuation}(a).

In order to obtain the dotted gray curve in Fig.~\ref{fig:scalar_continuation}(a), we expand
eq.~\eqref{eq:scalar_small_v_potential} at small $\gamma$:
\begin{align}
F_2=-192\gamma x^5(1-x^2)+\calo(\gamma^2).
\end{align}
The minimum of $-x^5(1-x^2)$ occurs at $x^2=5/7$. Define
\begin{align*}
 C_0\equiv\left[192\,\frac{2}{7}
\left(\frac{5}{7}\right)^{5/2}\right]^{-1/2}.
\end{align*}
Expanding the full expression in eq.~\eqref{eq:scalar_small_v_potential} at this point gives%
\footnote{
The location $x_{\rm  min}(\gamma)$ of the minimum of $F_2(x)$ shifts from $x=\sqrt{5/7}$ by $O(\gamma)$, however this shift first changes the minimum value of $F_2$ at $O(\gamma^3)$. 
}
\begin{align*}
\min F_2=-\frac{\gamma}{C_0^2}
\left[1-\frac{28260}{343}\gamma+O(\gamma^2)\right].
\end{align*}
Taking the inverse square root therefore gives
\begin{align}
D_2(\gamma)
&=\frac{C_0}{\sqrt{\gamma}}
\left[1+\frac{14130}{343}\gamma+\calo(\gamma^2)\right].
\label{eq:scalar_small_v_small_gamma}
\end{align}

\subsection{The limit $v\to1$}
\label{app:scalar_v_one}

\subsubsection{Fixed-$\gamma$ asymptotics}
Let $E_0(\hbar)$ denote the lowest eigenvalue of the generalized Schr\"odinger problem~\eqref{eq:scalar_seed_generalized_spectrum} at fixed $\gamma$. As $v\to1$, the ground state energy $E_0=-1/v^2$ approaches $-1$. Numerical evidence from Sec.~\ref{sec:scalar_seed} indicates that $\Gamma\to0$ as $v\to1$, and from eq.~\eqref{eq:scalar_seed_dictionary} we have $\Gamma = \sqrt{1-v^2}/(\hbar v)$. While $\hbar \to 0$ in the limit of $v\to1$ is consistent with vanishing $\Gamma$, we know from section~\ref{app:scalar_small_v} that $\hbar\to0$ gives $E\to-\infty$, rather than $E\to-1$. Moreover, numerical evidence from Sec.~\ref{sec:scalar_complex_momentum} indicates that $E=-1$ (threshold of causality violation) is achieved at a finite value of $|\q|$, which we call $\kappa_c$. We thus define $\hbar_c$ by%
\footnote{
  Thus, $\kappa_c$ is the magnitude of $\q$ at which the magenta curve in Fig.~\ref{fig:scalar_complex_momentum}(b) becomes causality-violating.
}
\begin{align}
E_0(\hbar_c)=-1,\qquad \kappa_c=\frac{1}{\hbar_c}\,,
\label{eq:scalar_v_one_threshold}
\end{align}
such that $\hbar_c$ is finite. Equation~\eqref{eq:scalar_seed_dictionary} then immediately gives the leading behavior
\begin{align}
  \Gamma(v)=\kappa_c\sqrt{1-v^2}\left[1+O(1{-}v^2)\right].
\label{eq:Gamma_v_kappa_c}
\end{align}
The $O(1{-}v^2)$ correction can also be determined. The state with $E=-1$ is a normalizable bound state (in the spectrum of $E<0$ states) with an isolated eigenvalue in the Schr\"odinger-equation problem. With $\delta\equiv1-v^2$, the ground-state energy is $E_0=-1-\delta+\dots$.
Alternatively, $E_0$ can be expanded as $E_0 = E_0(\hbar_c) + (\hbar - \hbar_c) E_0'(\hbar_c) + \dots$, thus
\begin{align}
  \hbar=\hbar_c-\frac{\delta}{E_0'(\hbar_c)}+ O(\delta^2)\,,
\end{align}
assuming $E_0'(\hbar_c) \neq0$.
Substituting into eq.~\eqref{eq:scalar_seed_dictionary}, we obtain
\begin{align}
\Gamma(v)= \frac{1}{\hbar_c} \sqrt{\delta}
\left[1+\left(\frac12+
\frac{1}{\hbar_c E_0'(\hbar_c)}\right)\delta
+ O(\delta^2)\right].
\label{eq:scalar_v_one_expansion}
\end{align}
In particular, the growth rate vanishes as $\sqrt{1-v^2}$, while $\operatorname{Im}\w$ and $|\q|$ approach the common finite value $\kappa_c$. This is the magenta-line crossing identified in Fig.~\ref{fig:scalar_complex_momentum}(b).

The derivative $E_0'(\hbar_c)$ (at fixed $\gamma$) can be evaluated by the Hellmann–Feynman theorem. Denoting the left-hand side of eq.~\eqref{eq:scalar_seed_generalized_spectrum} as ${\cal H}\psi_2$, we have $E'(\hbar) = \langle \psi_2 | {\cal H}'(\hbar) |\psi_2 \rangle / \langle \psi_2 | R_2 |\psi_2\rangle $ (recall that $R_2$ is independent of $\hbar$). Explicitly,
\begin{align}
  E_0'(\hbar_c)=
\frac{\displaystyle
2\hbar_c\int_{-\infty}^{0}|\partial_y\psi_c|^2dy+
\int_{-\infty}^{0}
\left.\partial_\hbar V_2\right|_{\hbar_c}|\psi_c|^2dy}
{\displaystyle\int_{-\infty}^{0}R_2|\psi_c|^2dy} \,,
\label{eq:scalar_v_one_derivative}
\end{align}
where $\psi_c$ is the state corresponding to $\kappa_c$, and the derivative of $V_2$ with respect to $\hbar$ includes contributions from both $U_2^{(-2)}$ and $U_2^{(2)}$. 
Numerically solving eq.~\eqref{eq:Scheq} at $\gamma=1/500$ to find $\kappa_c$ and $\psi_c$, and evaluating the integral in eq.~\eqref{eq:scalar_v_one_derivative} using the resulting $\psi_c$, one finds
\begin{align}
\hbar_c\simeq 0.206(0),\qquad
E_0'(\hbar_c)\simeq 7.335(7).
\end{align}
Consequently, at $\gamma=1/500$, eq.~\eqref{eq:scalar_v_one_expansion} gives
\begin{align}
\Gamma(v)= 4.854(5) \sqrt{1-v^2}
\left[1+ 1.161(8) (1-v^2)
+ O\bigl((1{-}v^2)^2\bigr)\right].
\label{eq:scalar_v_one_numerical}
\end{align}
The leading term in eq.~\eqref{eq:scalar_v_one_numerical} gives the dashed black curve in Fig.~\ref{fig:scalar_continuation}(b).

\subsubsection{Small-$\gamma$ expansion}
We now ask how $\kappa_c$ which controls the leading order of the small-$(1{-}v^2)$ expansion of $\Gamma(v)$ in \eqref{eq:Gamma_v_kappa_c} depends on $\gamma$. The relation between $\kappa_c$ and $\gamma$ is determined by the Schr\"odinger equation \eqref{eq:Scheq} with fixed eigenvalue $E=-1$. We are interested in $\kappa_c(\gamma)$ at small $\gamma$. Let $\sigma\equiv\gamma/\hbar^2 = \gamma \kappa^2$. We know from sec.~\ref{sec:universal_bound_states} that the potential in the Schr\"odinger equation \eqref{eq:Scheq} has a finite non-trivial limit as $\gamma\to0$, $\hbar\to0$, with $\sigma$ fixed. Motivated by that, we write $\sigma_c = \gamma \kappa_c^2$, and aim to find $\sigma_c(\gamma)$, which then determines $\kappa_c(\gamma) = \sqrt{\sigma_c(\gamma)/\gamma}$.
 
In fact, the first two terms in the small-$\gamma$ expansion of $\kappa_c(\gamma)$ can be read off directly from the results of sec.~\ref{sec:universal_bound_states}. Indeed, eq.~\eqref{eq:universal_quantum_threshold} gives
\begin{align}
  \gamma_c(\hbar) = \sigma_0 \hbar^2 (1 + a \hbar + O(\hbar^2))\,,
\end{align}
where $\sigma_0 = 25\sqrt{15}/3456$, and $a = \sqrt{6\sqrt{15}}/3$. Equivalently, $\sigma_c = \sigma_0 + \sigma_1 \sqrt{\gamma} + O(\gamma)$, where $\sigma_1 = a \sqrt{\sigma_0} = 5\sqrt{15}/72$. Expanding $\kappa_c = \sqrt{\sigma_c /\gamma}$ gives
\begin{align}
\kappa_c(\gamma)=\frac{15^{1/4}}{\sqrt6}
\left[\frac{5}{24\sqrt\gamma}+1 + O(\sqrt{\gamma})\right].
\label{eq:scalar_v_one_small_gamma-0}
\end{align}
Further terms in the small-$\gamma$ expansion of $\kappa_c(\gamma)$ may be obtained by taking into account the first $O(\gamma)$ correction to the potential in the scaling limit \eqref{eq:c-limit}, and by expanding the potential beyond the harmonic-oscillator approximation of sec.~\ref{sec:universal_bound_states}. We omit the details, and quote the next correction: 
\begin{align}
\sigma_c=\frac{25\sqrt{15}}{3456}
+\frac{5\sqrt{15}}{72}\sqrt{\gamma}
+\frac{14323\sqrt{15}}{25920}\gamma
+\calo(\gamma^{3/2}).
\end{align}
Substituting into $\kappa_c=\sqrt{\sigma_c/\gamma}$, this yields
\begin{align}
\kappa_c(\gamma)=\frac{15^{1/4}}{\sqrt6}
\left[\frac{5}{24\sqrt\gamma}+1+
\frac{10003}{1800}\sqrt\gamma+\calo(\gamma)\right].
\label{eq:scalar_v_one_small_gamma}
\end{align}

\section{Helicity-2 equation in the order-$\gamma^2$ extension}
\label{app:gamma2_coeffs}
In the equations below, $E = -\w^2/\q^2$, $\hbar^2 = -1/\q^2$, where $\w$ and $\q$ are defined by eq.~\eqref{eq:gamma2_frequency_momentum}, with the temperature in eq.~\eqref{eq:gamma2_temperature} and $r_0=1/2$. 

The helicity-2 variable used in this five-dimensional model differs from the ten-dimensional variable in Appendix~\ref{app:QNM_coeffs_scalar} by a regular radial normalization. Through order $\gamma$, the two variables are related by
\begin{align*}
 Z_2^{(5d)}=
\left[1+\frac{25}{16}\gamma x^4(1+x^2)\right]
Z_2^{(10d)}+\calo(\gamma^2).
\end{align*}
 This redefinition accounts for the different order-$\gamma$ coefficients in the two appendices; the corresponding Schr\"odinger potentials agree through that order.

For the phenomenological model \eqref{eq:gamma2_action}, the
order-reduced helicity-2 equation \eqref{eq:gamma2_reduced} can be
written compactly as follows. Define
\begin{align}
K\equiv4\pi^2T^2,\qquad
f\equiv1-x^2,\qquad
a(x)\equiv-\frac{1+x^2}{xf}.
\end{align}
Then
\begin{align}
\widetilde{\cal A}_2
=a+H+\frac{J}{\hbar^2}+\frac{L}{\hbar^4},
\label{eq:gamma2_A}
\end{align}
where
\begin{align}
H={}&
6\gamma x(129x^4+94x^2-25)
+\frac{18}{7}\gamma^2xP,
\notag\\
J={}&
-128K\gamma x^4
\left(\frac{15}{2}+\gamma Q\right),
\notag\\
L={}&
-307200K^2\gamma^2x^9.
\label{eq:gamma2_HJL}
\end{align}
The coefficient of $Z_2$ is
\begin{align}
\widetilde{\cal B}_2
={}&
\frac{K}{\hbar^2xf^2}
\Biggl\{
E+f
+\gamma x^2
\Bigl[
6E(89x^4-119x^2+25)
+f(851x^4-789x^2+75)
\Bigr]
\notag\\
&\hspace{28mm}
+\frac{\gamma^2x^2}{7}(EX+fY)
\Biggr\}
\notag\\
&+
\frac{192K^2x^4}{\hbar^4f}
(\gamma+\gamma^2S)
+\frac{36864K^3\gamma^2x^9}{\hbar^6f}.
\label{eq:gamma2_B}
\end{align}
The polynomials entering these expressions are
\begin{align}
P={}&
\alpha\Bigl(
894740x^{12}+875400x^{10}-110500x^8-88400x^6
-66300x^4-44200x^2-22100
\Bigr)
\notag\\
&+
8415099x^{10}+1721125x^8-4519466x^6
-115605x^4-93695x^2-69860,
\\
Q={}&
7020\alpha x^8+89342x^6
-(66204+3888E)x^4+1125,
\\
S={}&
360\alpha x^8+10687x^6
-(9245+3904E)x^4+75x^2+150,
\\
X={}&
15661800\alpha x^{12}-17746200\alpha x^{10}
+397800\alpha(x^8+x^6+x^4+x^2+1)
\notag\\
&+
104522733x^{10}-185635926x^8+77900823x^6
+1191330x^4+468405x^2+1257480,
\\
Y={}&
17476020\alpha x^{12}-17945100\alpha x^{10}
+198900\alpha(x^8+x^6+x^4+x^2+1)
\notag\\
&+
135086623x^{10}-213413970x^8+77651133x^6
+1476090x^4-239085x^2+628740.
\label{eq:gamma2_polynomials}
\end{align}
Primes in constructing the Schr\"odinger potential denote
$x$ derivatives, with $E,T,\alpha$ and $\gamma$ held fixed.
The expressions above are perturbatively equivalent to the
higher-derivative helicity-2 fluctuation equation through
order $\gamma^2$.

\section{Helicity-2 potential in the order-$\gamma^2$ extension}
\label{app:gamma2_potential}

For the extended coefficients of Appendix~\ref{app:gamma2_coeffs},
the Schr\"odinger potential has the decomposition
\begin{align}
\widetilde U_2
={}&
\hbar^{-6}\widetilde U_2^{(-6)}
+\hbar^{-4}\widetilde U_2^{(-4)}
+\hbar^{-2}\widetilde U_2^{(-2)}
+\widetilde U_2^{(0)}
+\hbar^2\widetilde U_2^{(2)}.
\end{align}
The leading terms of its components, keeping $T$ explicit, are
\begin{align}
\widetilde U_2^{(-6)}
={}&
6039797760000\,
\pi^8T^8\gamma^4x^{19}(1-x^2)^2,
\label{eq:gamma2_Um6}
\\
\widetilde U_2^{(-4)}
={}&
-2359296\,
\pi^6T^6\gamma^2x^{10}(1-x^2)
+\calo(\gamma^3),
\label{eq:gamma2_Um4}
\\
\widetilde U_2^{(-2)}
={}&
-3072\pi^4T^4\gamma x^5(1-x^2)
+\calo(\gamma^2),
\label{eq:gamma2_Um2}
\\
\widetilde U_2^{(0)}
={}&
(1-4\pi^2T^2)E-4\pi^2T^2(1-x^2)
\notag\\
&-4\pi^2T^2\gamma x^2
\Bigl[
6(89x^4-119x^2+25)E
+1549x^6-2200x^4+576x^2+75
\Bigr]
+\calo(\gamma^2),
\label{eq:gamma2_U0}
\\
\widetilde U_2^{(2)}
={}&
\frac{3(1-x^2)(3x^2+5)}{16x}
-6\gamma x^3(1-x^2)(387x^4-70x^2-119)
+\calo(\gamma^2).
\label{eq:gamma2_U2}
\end{align}
Equation~\eqref{eq:gamma2_Um6} is exact within the specified
resummed second-order equation. Its order-$\gamma^4$ origin is a
consequence of the nonlinear appearance of
$\widetilde{\cal A}_2$ in the Liouville transformation and does
not represent the complete order-$\gamma^4$ effective action.

Equations~\eqref{eq:gamma2_Um2}--\eqref{eq:gamma2_U2} agree
exactly through order $\gamma$ with the ten-dimensional
helicity-2 expressions in Appendix~\ref{ap:us}, including the
explicit temperature dependence. This comparison has been
performed only in the helicity-2 channel. The additional
$\widetilde U_2^{(-4)}$ and $\widetilde U_2^{(-6)}$ components
are generated by the order-$\gamma^2$ reduced equation and its
specified resummation, respectively, and were absent in the
order-$\gamma$ equation used in the main analysis.

\bibliographystyle{JHEP}
\bibliography{ads-gb}{}

\providecommand{\href}[2]{#2}\begingroup\raggedright\begin{thebibliography}{10}

\bibitem{Ammon:2015wua}
M.~Ammon and J.~Erdmenger, \emph{{Gauge/gravity duality}: {Foundations and
  applications}}, Cambridge University Press (2015),
  \href{https://doi.org/10.1017/CBO9780511846373}{10.1017/CBO9780511846373}.

\bibitem{Gubser:1998nz}
S.S.~Gubser, I.R.~Klebanov and A.A.~Tseytlin, \emph{{Coupling constant
  dependence in the thermodynamics of N=4 supersymmetric Yang-Mills theory}},
  \href{https://doi.org/10.1016/S0550-3213(98)00514-8}{\emph{Nucl. Phys. B}
  {\bfseries 534} (1998) 202}
  [\href{https://arxiv.org/abs/hep-th/9805156}{{\ttfamily hep-th/9805156}}].

\bibitem{Buchel:2004di}
A.~Buchel, J.T.~Liu and A.O.~Starinets, \emph{{Coupling constant dependence of
  the shear viscosity in N=4 supersymmetric Yang-Mills theory}},
  \href{https://doi.org/10.1016/j.nuclphysb.2004.11.055}{\emph{Nucl. Phys. B}
  {\bfseries 707} (2005) 56}
  [\href{https://arxiv.org/abs/hep-th/0406264}{{\ttfamily hep-th/0406264}}].

\bibitem{Benincasa:2005qc}
P.~Benincasa and A.~Buchel, \emph{{Transport properties of N=4 supersymmetric
  Yang-Mills theory at finite coupling}},
  \href{https://doi.org/10.1088/1126-6708/2006/01/103}{\emph{JHEP} {\bfseries
  01} (2006) 103} [\href{https://arxiv.org/abs/hep-th/0510041}{{\ttfamily
  hep-th/0510041}}].

\bibitem{Buchel:2008sh}
A.~Buchel, \emph{{Resolving disagreement for $\eta/s$ in a CFT plasma at finite
  coupling}},
  \href{https://doi.org/10.1016/j.nuclphysb.2008.05.024}{\emph{Nucl. Phys. B}
  {\bfseries 803} (2008) 166}
  [\href{https://arxiv.org/abs/0805.2683}{{\ttfamily 0805.2683}}].

\bibitem{Myers:2008yi}
R.C.~Myers, M.F.~Paulos and A.~Sinha, \emph{{Quantum corrections to eta/s}},
  \href{https://doi.org/10.1103/PhysRevD.79.041901}{\emph{Phys. Rev. D}
  {\bfseries 79} (2009) 041901}
  [\href{https://arxiv.org/abs/0806.2156}{{\ttfamily 0806.2156}}].

\bibitem{Stricker:2013lma}
S.A.~Stricker, \emph{{Holographic thermalization in N=4 Super Yang-Mills theory
  at finite coupling}},
  \href{https://doi.org/10.1140/epjc/s10052-014-2727-4}{\emph{Eur. Phys. J. C}
  {\bfseries 74} (2014) 2727}
  [\href{https://arxiv.org/abs/1307.2736}{{\ttfamily 1307.2736}}].

\bibitem{Waeber:2015oka}
S.~Waeber, A.~Sch{\"a}fer, A.~Vuorinen and L.G.~Yaffe, \emph{{Finite coupling
  corrections to holographic predictions for hot QCD}},
  \href{https://doi.org/10.1007/JHEP11(2015)087}{\emph{JHEP} {\bfseries 11}
  (2015) 087} [\href{https://arxiv.org/abs/1509.02983}{{\ttfamily
  1509.02983}}].

\bibitem{Grozdanov:2016vgg}
S.~Grozdanov, N.~Kaplis and A.O.~Starinets, \emph{{From strong to weak coupling
  in holographic models of thermalization}},
  \href{https://doi.org/10.1007/JHEP07(2016)151}{\emph{JHEP} {\bfseries 07}
  (2016) 151} [\href{https://arxiv.org/abs/1605.02173}{{\ttfamily
  1605.02173}}].

\bibitem{Casalderrey-Solana:2018rle}
J.~Casalderrey-Solana, S.~Grozdanov and A.O.~Starinets, \emph{{Transport Peak
  in the Thermal Spectral Function of $\mathcal{N}=4$ Supersymmetric Yang-Mills
  Plasma at Intermediate Coupling}},
  \href{https://doi.org/10.1103/PhysRevLett.121.191603}{\emph{Phys. Rev. Lett.}
  {\bfseries 121} (2018) 191603}
  [\href{https://arxiv.org/abs/1806.10997}{{\ttfamily 1806.10997}}].

\bibitem{Grozdanov:2018kkt}
S.~Grozdanov, \emph{{On the connection between hydrodynamics and quantum chaos
  in holographic theories with stringy corrections}},
  \href{https://doi.org/10.1007/JHEP01(2019)048}{\emph{JHEP} {\bfseries 01}
  (2019) 048} [\href{https://arxiv.org/abs/1811.09641}{{\ttfamily
  1811.09641}}].

\bibitem{Grozdanov:2021jfw}
S.~Grozdanov, A.O.~Starinets and P.~Tadi{\'c}, \emph{{Hydrodynamic dispersion
  relations at finite coupling}},
  \href{https://doi.org/10.1007/JHEP06(2021)180}{\emph{JHEP} {\bfseries 06}
  (2021) 180} [\href{https://arxiv.org/abs/2104.11035}{{\ttfamily
  2104.11035}}].

\bibitem{Buchel:2018eax}
A.~Buchel, \emph{{Sensitivity of holographic $\mathcal{N}=4$ SYM plasma
  hydrodynamics to finite coupling corrections}},
  \href{https://doi.org/10.1103/PhysRevD.98.061901}{\emph{Phys. Rev. D}
  {\bfseries 98} (2018) 061901}
  [\href{https://arxiv.org/abs/1807.05457}{{\ttfamily 1807.05457}}].

\bibitem{Buchel:2026rep}
A.~Buchel, R.E.~Hoult and P.~Kovtun, \emph{{Instability of 5D Gauss-Bonnet
  black branes}}, \href{https://doi.org/10.1007/JHEP10(2026)052}{\emph{JHEP}
  {\bfseries 10} (2026) 052}
  [\href{https://arxiv.org/abs/2606.19049}{{\ttfamily 2606.19049}}].

\bibitem{Brigante:2007nu}
M.~Brigante, H.~Liu, R.C.~Myers, S.~Shenker and S.~Yaida, \emph{{Viscosity
  Bound Violation in Higher Derivative Gravity}},
  \href{https://doi.org/10.1103/PhysRevD.77.126006}{\emph{Phys. Rev. D}
  {\bfseries 77} (2008) 126006}
  [\href{https://arxiv.org/abs/0712.0805}{{\ttfamily 0712.0805}}].

\bibitem{Brigante:2008gz}
M.~Brigante, H.~Liu, R.C.~Myers, S.~Shenker and S.~Yaida, \emph{{The Viscosity
  Bound and Causality Violation}},
  \href{https://doi.org/10.1103/PhysRevLett.100.191601}{\emph{Phys. Rev. Lett.}
  {\bfseries 100} (2008) 191601}
  [\href{https://arxiv.org/abs/0802.3318}{{\ttfamily 0802.3318}}].

\bibitem{Buchel:2009tt}
A.~Buchel and R.C.~Myers, \emph{{Causality of Holographic Hydrodynamics}},
  \href{https://doi.org/10.1088/1126-6708/2009/08/016}{\emph{JHEP} {\bfseries
  08} (2009) 016} [\href{https://arxiv.org/abs/0906.2922}{{\ttfamily
  0906.2922}}].

\bibitem{Camanho:2014apa}
X.O.~Camanho, J.D.~Edelstein, J.~Maldacena and A.~Zhiboedov, \emph{{Causality
  Constraints on Corrections to the Graviton Three-Point Coupling}},
  \href{https://doi.org/10.1007/JHEP02(2016)020}{\emph{JHEP} {\bfseries 02}
  (2016) 020} [\href{https://arxiv.org/abs/1407.5597}{{\ttfamily 1407.5597}}].

\bibitem{Grisaru:1986vi}
M.T.~Grisaru and D.~Zanon, \emph{{$\sigma$ Model Superstring Corrections to the
  Einstein-Hilbert Action}},
  \href{https://doi.org/10.1016/0370-2693(86)90765-3}{\emph{Phys. Lett. B}
  {\bfseries 177} (1986) 347}.

\bibitem{Freeman:1986zh}
M.D.~Freeman, C.N.~Pope, M.F.~Sohnius and K.S.~Stelle, \emph{{Higher Order
  $\sigma$ Model Counterterms and the Effective Action for Superstrings}},
  \href{https://doi.org/10.1016/0370-2693(86)91495-4}{\emph{Phys. Lett. B}
  {\bfseries 178} (1986) 199}.

\bibitem{Park:1987jp}
Q.-H.~Park and D.~Zanon, \emph{{More on $\sigma$ Model Beta Functions and
  Low-energy Effective Actions}},
  \href{https://doi.org/10.1103/PhysRevD.35.4038}{\emph{Phys. Rev. D}
  {\bfseries 35} (1987) 4038}.

\bibitem{Gross:1986iv}
D.J.~Gross and E.~Witten, \emph{{Superstring Modifications of Einstein's
  Equations}}, \href{https://doi.org/10.1016/0550-3213(86)90429-3}{\emph{Nucl.
  Phys. B} {\bfseries 277} (1986) 1}.

\bibitem{Buchel:2008ac}
A.~Buchel, \emph{{Shear viscosity of boost invariant plasma at finite
  coupling}},
  \href{https://doi.org/10.1016/j.nuclphysb.2008.03.009}{\emph{Nucl. Phys. B}
  {\bfseries 802} (2008) 281}
  [\href{https://arxiv.org/abs/0801.4421}{{\ttfamily 0801.4421}}].

\bibitem{Buchel:2008bz}
A.~Buchel and M.~Paulos, \emph{{Relaxation time of a CFT plasma at finite
  coupling}},
  \href{https://doi.org/10.1016/j.nuclphysb.2008.07.002}{\emph{Nucl. Phys. B}
  {\bfseries 805} (2008) 59} [\href{https://arxiv.org/abs/0806.0788}{{\ttfamily
  0806.0788}}].

\bibitem{Buchel:2008kd}
A.~Buchel and M.~Paulos, \emph{{Second order hydrodynamics of a CFT plasma from
  boost invariant expansion}},
  \href{https://doi.org/10.1016/j.nuclphysb.2008.10.012}{\emph{Nucl. Phys. B}
  {\bfseries 810} (2009) 40} [\href{https://arxiv.org/abs/0808.1601}{{\ttfamily
  0808.1601}}].

\bibitem{Paulos:2008tn}
M.F.~Paulos, \emph{{Higher derivative terms including the Ramond-Ramond
  five-form}}, \href{https://doi.org/10.1088/1126-6708/2008/10/047}{\emph{JHEP}
  {\bfseries 10} (2008) 047} [\href{https://arxiv.org/abs/0804.0763}{{\ttfamily
  0804.0763}}].

\bibitem{Green:1999pu}
M.B.~Green, H.-h.~Kwon and P.~Vanhove, \emph{{Two loops in eleven-dimensions}},
  \href{https://doi.org/10.1103/PhysRevD.61.104010}{\emph{Phys. Rev. D}
  {\bfseries 61} (2000) 104010}
  [\href{https://arxiv.org/abs/hep-th/9910055}{{\ttfamily hep-th/9910055}}].

\bibitem{Aharony:1999ti}
O.~Aharony, S.S.~Gubser, J.M.~Maldacena, H.~Ooguri and Y.~Oz, \emph{{Large N
  field theories, string theory and gravity}},
  \href{https://doi.org/10.1016/S0370-1573(99)00083-6}{\emph{Phys. Rept.}
  {\bfseries 323} (2000) 183}
  [\href{https://arxiv.org/abs/hep-th/9905111}{{\ttfamily hep-th/9905111}}].

\bibitem{Pawelczyk:1998pb}
J.~Pawelczyk and S.~Theisen, \emph{{$AdS_5 \times S^5$ black hole metric at
  $O(\alpha'^3)$}},
  \href{https://doi.org/10.1088/1126-6708/1998/09/010}{\emph{JHEP} {\bfseries
  09} (1998) 010} [\href{https://arxiv.org/abs/hep-th/9808126}{{\ttfamily
  hep-th/9808126}}].

\bibitem{Wald:1993nt}
R.M.~Wald, \emph{{Black hole entropy is the Noether charge}},
  \href{https://doi.org/10.1103/PhysRevD.48.R3427}{\emph{Phys. Rev. D}
  {\bfseries 48} (1993) R3427}
  [\href{https://arxiv.org/abs/gr-qc/9307038}{{\ttfamily gr-qc/9307038}}].

\bibitem{Dutta:2006vs}
S.~Dutta and R.~Gopakumar, \emph{{On Euclidean and Noetherian entropies in AdS
  space}}, \href{https://doi.org/10.1103/PhysRevD.74.044007}{\emph{Phys. Rev.
  D} {\bfseries 74} (2006) 044007}
  [\href{https://arxiv.org/abs/hep-th/0604070}{{\ttfamily hep-th/0604070}}].

\bibitem{Skenderis:2002wp}
K.~Skenderis, \emph{{Lecture notes on holographic renormalization}},
  \href{https://doi.org/10.1088/0264-9381/19/22/306}{\emph{Class. Quant. Grav.}
  {\bfseries 19} (2002) 5849}
  [\href{https://arxiv.org/abs/hep-th/0209067}{{\ttfamily hep-th/0209067}}].

\bibitem{Kovtun:2005ev}
P.K.~Kovtun and A.O.~Starinets, \emph{{Quasinormal modes and holography}},
  \href{https://doi.org/10.1103/PhysRevD.72.086009}{\emph{Phys. Rev. D}
  {\bfseries 72} (2005) 086009}
  [\href{https://arxiv.org/abs/hep-th/0506184}{{\ttfamily hep-th/0506184}}].

\bibitem{Buchel:2009sk}
A.~Buchel, J.~Escobedo, R.C.~Myers, M.F.~Paulos, A.~Sinha and M.~Smolkin,
  \emph{{Holographic GB gravity in arbitrary dimensions}},
  \href{https://doi.org/10.1007/JHEP03(2010)111}{\emph{JHEP} {\bfseries 03}
  (2010) 111} [\href{https://arxiv.org/abs/0911.4257}{{\ttfamily 0911.4257}}].

\bibitem{Nunez:2003eq}
A.~Nunez and A.O.~Starinets, \emph{{AdS / CFT correspondence, quasinormal
  modes, and thermal correlators in N=4 SYM}},
  \href{https://doi.org/10.1103/PhysRevD.67.124013}{\emph{Phys. Rev. D}
  {\bfseries 67} (2003) 124013}
  [\href{https://arxiv.org/abs/hep-th/0302026}{{\ttfamily hep-th/0302026}}].

\bibitem{Festuccia:2008zx}
G.~Festuccia and H.~Liu, \emph{{A Bohr-Sommerfeld quantization formula for
  quasinormal frequencies of AdS black holes}},
  \href{https://doi.org/10.1166/asl.2009.1029}{\emph{Adv. Sci. Lett.}
  {\bfseries 2} (2009) 221} [\href{https://arxiv.org/abs/0811.1033}{{\ttfamily
  0811.1033}}].

\bibitem{Grozdanov:2019kge}
S.~Grozdanov, P.K.~Kovtun, A.O.~Starinets and P.~Tadi{\'c}, \emph{{Convergence
  of the Gradient Expansion in Hydrodynamics}},
  \href{https://doi.org/10.1103/PhysRevLett.122.251601}{\emph{Phys. Rev. Lett.}
  {\bfseries 122} (2019) 251601}
  [\href{https://arxiv.org/abs/1904.01018}{{\ttfamily 1904.01018}}].

\bibitem{Grozdanov:2019uhi}
S.~Grozdanov, P.K.~Kovtun, A.O.~Starinets and P.~Tadi{\'c}, \emph{{The complex
  life of hydrodynamic modes}},
  \href{https://doi.org/10.1007/JHEP11(2019)097}{\emph{JHEP} {\bfseries 11}
  (2019) 097} [\href{https://arxiv.org/abs/1904.12862}{{\ttfamily
  1904.12862}}].

\bibitem{Benincasa:2005iv}
P.~Benincasa, A.~Buchel and A.O.~Starinets, \emph{{Sound waves in strongly
  coupled non-conformal gauge theory plasma}},
  \href{https://doi.org/10.1016/j.nuclphysb.2005.11.005}{\emph{Nucl. Phys. B}
  {\bfseries 733} (2006) 160}
  [\href{https://arxiv.org/abs/hep-th/0507026}{{\ttfamily hep-th/0507026}}].

\end{thebibliography}\endgroup

\end{document}